\documentclass[a4paper,11pt]{article}
\usepackage{geometry}
\usepackage{a4wide}
\usepackage{graphicx}
\usepackage{epsf}
\usepackage{amsmath}
\usepackage{amssymb}
\usepackage[normalem]{ulem}

\usepackage{cite}
\usepackage{multirow}
\usepackage{subfigure}
\usepackage{appendix}
\usepackage{tikz}
\usepackage{braket}
\usepackage{amsfonts,amsthm,euscript,braket,xcolor}
\usepackage{breqn}
\usepackage[colorlinks=true, linkcolor=blue, citecolor=blue, urlcolor=blue]{hyperref}  

\newcommand{\be}{\begin{equation}}
	\newcommand{\ee}{\end{equation}}

\newcommand{\Rmnum}[1]{\expandafter\@slowromancap\romannumeral #1@}
\newcommand{\bea}{\begin{eqnarray}}
	\newcommand{\eea}{\end{eqnarray}}

\numberwithin{equation}{section}

\usepackage{ulem}

\begin{document}
	
	
	\title{\bf A dynamical Einstein-Born-Infeld-dilaton model and holographic quarkonium melting in a magnetic field
 }
	\author{ \textbf{\textbf{Siddhi Swarupa Jena$^{a}$}\thanks{519ph2015@nitrkl.ac.in}, \textbf{Jyotirmoy Barman$^{a,b}$}\thanks{jyotirmoy.barman@icts.res.in}, Bruno Toniato$^{c,d}$}\thanks{bruno.toniato@ufabc.edu.br}, \\ \textbf{David Dudal$^{c}$}\thanks{david.dudal@kuleuven.be}, \textbf{Subhash Mahapatra$^{a}$}\thanks{mahapatrasub@nitrkl.ac.in}
		\\\\\textit{{\small $^a$ Department of Physics and Astronomy, National Institute of Technology Rourkela,}}\\
  \textit{{\small Rourkela - 769008, India}}\\
  \textit{{\small $^b$ International Center for Theoretical Sciences, Bengaluru, Bengaluru-560089, India}}\\
		\textit{{\small $^c$ KU Leuven Campus Kortrijk--Kulak, Department of Physics, Etienne Sabbelaan 53 bus 7657,}}\\
		\textit{{\small 8500 Kortrijk, Belgium}}\\
            \textit{{\small $^d$  Centro de Ciências Naturais e 
        Humanas, Universidade Federal do ABC (UFABC),}}\\
            \textit{{\small 09210-170 Santo André, São Paulo,   
        Brazil}}
	}
	
	\date{}
\date{}
\maketitle
\begin{abstract}
	We generalize the potential reconstruction method to set up a dynamical Einstein-Born-Infeld-dilaton model, which we then use to study holographic quarkonium melting in an external magnetic field. The non-linear nature of the model allows to couple the magnetic field to the quarkonium inner structure without having to introduce back-reacting charged flavour degrees of freedom. The magnetic field dependent melting temperature is computed from the spectral functions and suggests a switch from inverse magnetic to magnetic catalysis when the magnetic field increases. We also discuss the differences due to the anisotropy brought in by the external field.
\end{abstract}

\section{Introduction}
\label{sec1}
Quantum chromodynamics (QCD) is deep-rooted as the quantum field theory of strong interactions, correctly elucidating the behavior of sub-atomic particles such as quarks and gluons and their interactions. At low temperatures and chemical potentials, QCD exhibits confinement and chiral symmetry breaking. Conversely, as the temperature and chemical potential increase, there is a transition to a phase characterized by the restoration of chiral symmetry. In this higher-temperature and chemical potential regime, deconfinement also emerges as a significant feature and a new exotic state of QCD matter is formed, the quark-gluon plasma (QGP). But it is very short-lived in earth circumstances, this plasma state is imagined to have existed straight away after the Big Bang, when the temperature was very high. Now it is only available after ultra-relativistic heavy ion collisions in controlled laboratory environments like the \textbf{LHC} Alice facility or \textbf{RHIC}. The QGP not only has remarkably different properties from the standard plasma \cite{SHURYAK200948,Apolin_rio_2022} but is also strongly coupled near the deconfinement temperature. The latter outcome can be deduced from the fact that the deconfinement temperature is of the same order as the fundamental QCD scale ($\Lambda_{QCD}$) at which QCD remains strongly interacting. This makes a common perturbative analysis intractable in the relevant and interesting parameter space of the QGP. Hence, in the vicinity of the deconfinement phase transition, the prevailing physics primarily manifests in a non-perturbative manner.

Recently, a potential extra parameter has been suggested that might non-trivially leave its footprints in the phase structure of QCD and may lead to a series of new strange QCD phenomena \cite{Miransky:2015ava,Kharzeev:2012ph,DElia:2010abb,Kharzeev:2007jp, Skokov:2009qp}. Indeed, a very strong magnetic field ($\sim10^{14}$ Tesla) is expected to be produced, though short-lived, during non-central heavy ion collisions \cite{Skokov:2009qp,Bzdak2012Mar,Voronyuk2011May,Deng2012Apr,Tuchin2022}. The notably large produced magnetic field quickly decays after the collision. However, it is predicted to be still fairly large enough in the course of time when QGP forms \cite{Tuchin2013Aug,McLerran2014Sep} or near the deconfinement phase transition temperature \cite{D'Elia2010Sep}. Therefore, a robust background magnetic field is expected to yield significant impacts on observables within QCD. This expectation has led to an intense investigation of magnetized QCD. Indeed, because of its phenomenological relevance for e.g.~the
chiral magnetic effect \cite{Fukushima:2008xe,Kharzeev:2007jp}, (inverse) magnetic catalysis \cite{Miransky:2002rp,Gatto:2010pt,Mizher:2010zb,Osipov:2007je,Kashiwa:2011js,Alexandre:2000yf,Fraga:2008um,Fukushima:2012xw,Semenoff:1999xv,Bali:2011qj,Bali:2012zg,Ilgenfritz:2013ara,Bruckmann:2013oba,Fukushima:2012kc,Ferreira:2014kpa,Mueller:2015fka,Bali:2013esa,Fraga:2012fs,Fraga:2012ev,Shovkovy:2012zn},
early universe physics \cite{Grasso:2000wj,Vachaspati:1991nm}, etc., magnetised QCD has drawn much attention in recent years. For comprehensive reviews on these subjects, see \cite{Miransky:2015ava,Kharzeev:2012ph,Endrodi:2024cqn}.

The initial works on magnetised QCD gave an incoherent and somewhat puzzling picture concerning the impact of the background magnetic field on the deconfinement and chiral transition temperature. In \cite{Gusynin1995Apr,Gusynin2022,Mizher:2010zb}, based on the linear sigma and Polyakov-Nambu-Jona-Lasinio models, an increase in the deconfinement transition temperature $T_c$ with the magnetic field was predicted, a phenomenon termed magnetic catalysis. On the other hand, an investigation of the non-linear sigma model predicted a decrease of $T_c$ with $B$ \cite{Agasian2008Jun}, that is,  inverse magnetic catalysis. The initial quenched lattice QCD investigations hinted towards inverse magnetic catalysis behavior \cite{DElia:2010abb}. The concerned picture becomes more or less clear after the availability of unquenched lattice QCD results (with full quark dynamics) which predicted inverse catalysis behavior \cite{Bali:2011qj}. Let us also refer to \cite{Ilgenfritz:2013ara,Ilgenfritz:2012fw,Mueller:2015fka,Ayala:2014iba}
for more work in these directions. 

In the present paper, we are interested in seeing the repercussions of the background magnetic field in the deconfinement phase transition and on the heavy quark bound state melting. Since the fundamental physics is non-perturbative near the deconfinement phase transition, we will use the gauge/gravity duality to model strongly coupled QCD in a magnetic field background \cite{Maldacena1999Apr,Witten:1998qj,Gubser:1998bc}. For this purpose, one can consider both top-down as well as bottom-up magnetised AdS/QCD models. The top-down models, though are consistently rooted in higher-dimensional string theory, and therefore are on solid footing as far as the validness of the duality is concerned, unfortunately, do not exhibit many important QCD features. For instance, if we consider the magnetised AdS background of \cite{DHoker:2009mmn,DHoker:2009ixq} to model QCD, we may not only face limitations in defining the confinement/deconfinement transition temperature,\footnote{The Einstein-Maxwell magnetised AdS background of \cite{DHoker:2009mmn,DHoker:2009ixq} does not exhibit a Hawking/Page type phase transition between the black hole and thermal-AdS. Hence, there is no confinement/deconfinement phase transition in the dual boundary theory. This problem can however be bypassed by introducing an additional dilaton field, without having a kinetic term, in the action, i.e., using a soft-wall type model phenomenology \cite{Dudal:2015wfn}. Unfortunately, these models are not entirely satisfactory in a way that they do not solve all the gravity field equations consistently.} but also face problems in modeling the running coupling constant, although see the recent \cite{Arefeva:2024vom}.  The bottom-up phenomenological models, on the other hand, are usually designed in an ad-hoc way after certain tweaks in the original models of the duality (without necessarily having much string theory rooting) and do exhibit many desirable magnetised QCD features \cite{Bohra:2019ebj,Bohra:2020qom,Dudal:2021jav,Gursoy:2017wzz,Gursoy:2016ofp,McInnes:2015kec,Arefeva:2020vae,Rougemont:2015oea,Finazzo:2016mhm}. More discussion on magnetised QCD from holography can be found in \cite{Rougemont:2014efa,Fuini:2015hba,Cartwright:2019opv,Fukushima:2021got,Ballon-Bayona:2022uyy,Rodrigues:2017cha,Rodrigues:2017iqi,Arefeva:2023jjh,Arefeva:2021jpa,Jena:2022nzw,Shukla:2023pbp,Jain:2022hxl,Dudal:2016joz,Zhou:2022izh,Arefeva:2022avn,Deng:2021kyd,Chen:2021gop,Braga:2020hhs,Zhou:2020ssi,Ballon-Bayona:2020xtf,Zhao:2021ogc,Dudal:2018rki,Arefeva:2024xmg,Ammon:2020rvg} , see also \cite{Rougemont:2023gfz,Hoyos:2021uff,Jarvinen:2021jbd,Gursoy:2010fj,Mamo:2015dea} for detailed reviews on holographic QCD or \cite{Giataganas:2012zy,Giataganas:2017koz,Zhu:2024uwu,Gursoy:2018ydr,Chernicoff:2012bu, Wang:2024rim} for works on anisotropic holographic QCD models in general. Here, we will construct and use a consistent bottom-up phenomenological magnetised AdS/QCD model to study deconfinement transition and heavy quark melting in a magnetic background.

A famous smoking gun signal of the formation of QGP is the suppression of heavy quarkonia (like $J/\psi$), a bound state composed of a heavy quark and anti-quark pair \cite{Matsui1986Oct,Iwasaki:2021nrz,Mattioli:2024vmr}. Essentially, it is related to the high probability of quarkonium breaking apart into free quarks as it travels through the QGP medium. Two primary mechanisms elucidate the diminution of quarkonium, encompassing Hot Nuclear Matter (HNM) effects and Cold Nuclear Matter (CNM) effects \cite{Zhao2023Jun}. The former attributes the reduction to the presence of the QGP medium, with temperature and baryon chemical potential serving as key factors. On the other hand, CNM effects stem from factors beyond the QGP medium. These include parton distribution within nuclei, inelastic collisions between quarkonium and nucleons (nuclear absorption), interactions between quarkonium and co-moving particles resulting in the dissolution of the bound state (co-mover dissociation), parton energy loss during multiple scattering processes (energy loss), and the electromagnetic field generated by bystanders (electromagnetic effect). Let us also refer to the lattice paper \cite{Asakawa:2003re}, where it was observed that the quarkonia could survive the deconfinement transition at temperature $T_c$ and can exist in temperatures up to $1.6~T_c$.

Therefore studying holographic quarkonium melting (suppression) could be a promising way to probe interesting features of QGP near the deconfinement transition. We undertake such an analysis in the current paper and investigate the thermal and magnetic field effects on the melting of heavy quarkonium using gauge/gravity duality inspired QCD models. This has been investigated previously \cite{Dudal:2014jfa,Braga:2018zlu}, see also \cite{Peeters2006Nov,Fujita2009Aug,Ishii2014Apr,Ali-Akbari2014Jun,Ali-Akbari2013Mar}. Generally, for such investigations, some soft-wall type models are considered on the magnetised AdS geometry of \cite{DHoker:2009mmn,DHoker:2009ixq} to study the magnetic field dependent anisotropic charmonia melting. For instance, in \cite{Dudal:2014jfa} the Born-Infeld (BI)  version of the soft-wall model was considered in the background of \cite{DHoker:2009mmn,DHoker:2009ixq}. Similarly, in \cite{Braga:2018zlu} the Maxwell based soft-wall model in the background of \cite{DHoker:2009mmn,DHoker:2009ixq} was used and the membrane paradigm approach was adopted to study quarkonium dissociation \cite{Iqbal:2008by}. Unfortunately, the soft-wall models are plagued with many inconsistencies in effectively modeling QCD from holography. For instance, not only the dilaton field is put into the action in an ad-hoc way but also the area law of the Wilson loop is not always respected in soft-wall models. Moreover, in soft-wall models, the dilaton field does not come from a consistent gravity solution. 

Another limitation of the above-mentioned studies is that, as mentioned earlier, there is no Hawking/Page phase transition in the considered magnetised AdS backgrounds of \cite{DHoker:2009mmn,DHoker:2009ixq}. Since this transition corresponds to the confinement/deconfinement phase transition in dual field theory, one therefore can not consistently define the deconfinement transition temperature. This limitation creates another issue. Generally, quarks are added in a probe brane approximation where their back-reaction on the dual geometric background (confining or deconfining) is neglected \cite{Karch:2002sh}. This means that there is no direct quark effect on the transition temperature. To find how the melting of quarkonia is influenced by a magnetic field, one needs to find a way to couple the magnetic field to the substructure of the quark-antiquark bound state. This was done first in \cite{Dudal:2014jfa,Dudal:2018rki}, albeit in soft-wall context, by considering Born-Infeld type action which allowed to incorporate to some extent the coupling of the magnetic field to the charged quark constituents in the $J/\psi$ quarkonia.

In this work, we overcome some of the above-mentioned issues and study magnetic field effects on the melting of heavy quarkonium by upgrading the soft-wall model of \cite{Dudal:2014jfa} to a consistent phenomenological bottom-up magnetised holographic QCD model. In particular, we consider the magnetic field embedded Einstein-Born-Infeld-dilaton action on the gravity side and solve the Einstein, Born-Infeld, and dilaton equations of motion analytically using the potential reconstruction method \cite{Dudal:2017max,Mahapatra:2018gig,He:2013qq,Arefeva:2018hyo,Arefeva:2020byn,Alanen:2009xs,Mahapatra:2020wym,Priyadarshinee:2021rch,Priyadarshinee:2023cmi,Cai:2012xh,Daripa:2024ksg}. We obtain closed-form magnetised AdS solutions, which not only have a non-trivial and consistent profile for the dilaton field but also exhibit a magnetic field dependent Hawking/Page phase transition. We then investigate the melting of heavy quarkonium in the deconfinement phase by introducing gauge field fluctuations in the Born-Infeld sector in two different directions: parallel and transverse to the magnetic field. The quarkonium melting is then analysed by obtaining the spectral function using two approaches: the real-time holographic prescription and the membrane paradigm method. 
 
The structure of this paper is outlined as follows. In Section~\ref{sec2}, we introduce the Einstein-Born-Infeld-dilaton gravity setup and discuss the corresponding solutions. Section~\ref{sec3} is dedicated to the computation of the Born-Infeld parameter ``$b$". Section~\ref{sec4} gives a detailed description of black hole thermodynamics. In Section~\ref{sec5}, we set up the equation of motion of fluctuations. Section~\ref{sec6} encompasses numerical results of the spectral function using the real-time holographic prescription. In section~\ref{sec7}, we study the spectral function using the membrane paradigm approach. Finally, in Section~\ref{sec8}, we encapsulate our primary findings and provide an outlook for future research directions. Concluding the paper is an appendix addressing two equivalent formulations of the Born-Infeld Lagrangian. 
 
 \section{Potential reconstruction method and a dynamical Einstein-Born-Infeld-dilaton setup} 
 \label{sec2}
 To develop a consistent dynamical version of the soft-wall model of \cite{Dudal:2014jfa}, we need to formulate a gravity model incorporating a background magnetic and a dilaton field. For this purpose, we consider the five-dimensional Einstein-Born-Infeld-dilaton (EBID) action,
 \begin{align}
S_{EBI} = & -\frac{1}{16 \pi G_5} \int \mathrm{d^5}x  \biggl[\sqrt{-g} \ \biggl(R + \frac{f(\phi)}{(2 \pi \alpha')^2} \  -\frac{1}{2}\partial_{\mu}\phi \partial^{\mu}\phi -V(\phi)\biggr) \nonumber \\
& - \frac{1}{(2 \pi \alpha')^2}\sqrt{-\det\biggl(g_{\mu \nu}+ 2 \pi \alpha' F_{\mu \nu}\biggr)}\biggr],
\label{actionEBI}
\end{align}
where $G_5$ is the five-dimensional Newton's constant, $R$ is the Ricci scalar, $F_{\mu \nu}$ is the field strength tensor of the $U(1)$ gauge field, and $\phi$ is the dilaton field. The $U(1)$ gauge field will be used to introduce a constant background magnetic field $B$. The coupling between the $U(1)$ gauge field and dilaton field is represented by gauge kinetic function $f(\phi)$ and $V(\phi)$ is the dilaton potential, to be determined later on.

The string parameter $\alpha'$ is a new dimensional parameter and is related to the Born-Infeld parameter $b$ by the relation $b=1/(2 \pi \alpha')$. By substituting the value of $b$ and rearranging the preceding Eq.~(\ref{actionEBI}), the resultant expression yields the modified form of the EBID action\footnote{The equivalence between equations (\ref{actionEBI}) and (\ref{actionEBImodified}) is shown in Appendix \ref{sec10}. }
\begin{eqnarray}
 S_{EBI} =  -\frac{1}{16 \pi G_5}\int \mathrm{d^5}x \sqrt{-g} \left[R + f(\phi) b^2 \left(1-\sqrt{1+\frac{F_{\mu \nu}F^{\mu \nu}}{2 b^2}}\right) -\frac{1}{2}\partial_{\mu}\phi \partial^{\mu}\phi -V(\phi)\right].
 	\label{actionEBImodified}
 \end{eqnarray}
As the parameter $b \rightarrow \infty$, the action collapses to the Einstein-Maxwell-dilaton theory. For a nice review of Born-Infeld electrodynamics, we refer the readers to \cite{Alam2021Nov}.

By varying the EBID action (\ref{actionEBImodified}) with respect to metric tensor, gauge field, and dilaton field, we can get the Einstein, gauge field, and dilation equations of motion, respectively. The Einstein equation of motion is given by,
\begin{eqnarray}
R_{\mu \nu} - \frac{1}{2}g_{\mu \nu}R - T_{\mu \nu}=0\,,
 	\label{einsteineom}
 \end{eqnarray}
where,
\begin{eqnarray}
 &&T_{\mu\nu}\\&&=\frac{1}{2}\left(\partial_{\mu}\phi\partial_{\nu}\phi-\frac{1}{2}g_{\mu \nu}(\partial\phi)^2-g_{\mu \nu}V(\phi)\right)+\frac{f(\phi)}{2}\left[\frac{F_{\mu \rho} F_{\nu}^{\rho}}{\sqrt{1+\frac{F^2}{2 b^2}}}+b^2g_{\mu \nu}\left(1-\sqrt{1+\frac{F^2}{2b^2}}\right)\right].\nonumber
\end{eqnarray}
The gauge field equation of motion is expressed as
\begin{eqnarray}
\partial_{\mu}\left[\sqrt{-g}\frac{f(\phi)}{\sqrt{1+\frac{F^2}{2b^2}}} F^{\mu \nu}\right]=0\,.
\label{maxwelleom}
\end{eqnarray}
Similarly, the dilaton field equation of motion is
\begin{eqnarray}
	\partial_{\mu}\left[\sqrt{-g} \ \partial^\mu\phi\right]-\sqrt{-g} \ \left[\frac{\partial V}{\partial \phi}- \frac{\partial f}{\partial \phi} \  b^2 \left(1-\sqrt{1+\frac{F^2}{2 \ b^2}}\right)\right]=0\,.
	\label{dilatoneom}
\end{eqnarray}
To solve the field equations, we consider the following Ans\"atze for the metric, gauge field, and dilaton field:
\begin{eqnarray}
	& & ds^2=\frac{L^2 e^{2A(z)}}{z^2}\biggl[-g(z)dt^2 + \frac{dz^2}{g(z)} + dx_{1}^2+ e^{B^2 z^2} \biggl( dx_{2}^2 + dx_{3}^2 \biggr) \biggr]\,, \nonumber \\
	& & \phi=\phi(z), \ \  F_{\mu \nu}=B dx_{2}\wedge dx_{3}\,,
	\label{ansatze}
\end{eqnarray}
where $A(z)$ is an overall scale factor, which is needed to model in confinement properties holographically. $L$ is the AdS length scale, which we will set to one throughout this paper for convenience. The parameter $B$ denotes the five-dimensional magnetic field having dimension GeV. The magnetic field in the four-dimensional dual boundary theory has a dimension GeV$^2$ and is related to $B$ by a suitable $L$-rescaling. In this coordinate system, a background magnetic field is introduced in the $x_1$-direction. Because of this background magnetic field, the boundary
system no longer enjoys the $SO(3)$ invariance in spatial
coordinates $(x_1,x_2,x_3)$. It is straightforward to see that for the above Ans\"atze the gauge field equation of motion is trivially satisfied.\par
 
Plugging the Ans\"atze~(\ref{ansatze}) in Eq.~(\ref{einsteineom}), we get the following Einstein equations of motion,
\begin{equation}
     \biggl(-\frac{3}{z} + 2 B^2 z + 3A'(z)\biggr)g'(z)+g''(z)=0 \,, \label{einstein1}
\end{equation}
\begin{equation}
    \frac{2 B^2}{3} + \frac{2 B^4 z^2}{3} + \frac{2 A'(z)}{z}- A'(z)^2 + \frac{1}{6} \phi'(z)^2+A''(z)=0 \,,\label{einstein2}
\end{equation}
\begin{eqnarray}
& & g'(z) + \sqrt{1+\frac{B^2 e ^{-2 B^2 z^2-4 A(z)} z^4}{b^2}} \frac{b^2 e^{2 A(z)} z f(z)}{2 \bigl (b^2 e ^{2 B^2 z^2+4 A(z)}+B^2 z^4\bigr)} \nonumber\\
& &  + g(z) \biggl(-\frac{2}{z}+ 2B^2 z+ 3 A'(z)\biggr)=0 \,, \label{einstein3}
\end{eqnarray}
\begin{align}
    & -10 B^2 + \frac{12}{z^2} + 4 B^4 z^2 -\frac{b^2 e^{2 A(z) f(z)}}{z^2 g(z)}\biggl(1-\sqrt{1+\frac{b^2 e ^{-2 B^2 z^2-4 A(z)} z^4}{b^2}}\biggr)+ \frac{e^{2 A(z)V(z)}}{z^2 g(z)}  \nonumber \\
    &+  A'(z) \biggl(12 B^2 z+ 9 A'(z)+\frac{9 g'(z)}{2 g(z)}-\frac{18}{z}\biggr) + \frac{3 g'(z)}{g(z)}\biggl(B^2 z - \frac{3}{2 z}\biggr) + 3 A''(z)+ \frac{g''(z)}{2 g(z)}=0 \,. \label{einstein4} 
\end{align}
Similarly, the dilaton field equation of motion is
\begin{align}
    &\phi''(z)+\biggl(-\frac{3}{z} + 2 B^2 z + 3 A'(z) + \frac{g'(z)}{g(z)}\biggr)\phi'(z)+\frac{e^{2 A(z)}V'(z)}{z^2 g(z)} \nonumber \\ 
    & + \frac{b^2 e^{2 A(z) f'(z)}}{z^2 g(z)} \biggl(1-\sqrt{1+\frac{B^2 e^{-2 B^2 z^2 - 4A(z)}z^4}{b^2}}\biggr) =0 \,. \label{dilatoneq}
\end{align}
Interestingly, using the following boundary conditions
\begin{eqnarray}
&& g(0)=1 \ \ \text{and} \ \ g(z_h)=0, \nonumber \\
&& A(0) = 0 \,,
\label{boundaryconditions}
\end{eqnarray}
in combination with the potential reconstruction method \cite{Dudal:2017max,Mahapatra:2018gig,He:2013qq,Arefeva:2018hyo,Arefeva:2020byn,Alanen:2009xs,Mahapatra:2020wym,Priyadarshinee:2021rch,Priyadarshinee:2023cmi,Cai:2012xh,Daripa:2024ksg}, the Einstein, gauge field, and dilaton equations of motion can be solved completely and a closed form expressions of various geometric quantities can be obtained.
The solutions are given by the following expressions,
\begin{align}
A(z) & = -a z^2 \,, \label{asol} \\
g(z) & = 1-\frac{e^{z^2 \left(3 a-B^2\right)} \left(3 a z^2-B^2 z^2-1\right)+1}{e^{z_h^2 \left(3 a-B^2\right)} \left(3 a z_h^2-B^2 z_h^2-1\right)+1} \,, \label{gsol} \\
\phi(z) & = \frac{\left(9 a-B^2\right) \log \left(\sqrt{6 a^2-B^4} \sqrt{z^2 \left(6 a^2-B^4\right)+9 a-B^2}+6 a^2 z -B^4 z \right)}{\sqrt{6 a^2-B^4}} \nonumber\\
& \quad + z \sqrt{z^2 \left(6 a^2-B^4\right)+9 a-B^2} -\frac{\left(9 a-B^2\right) \log \left(\sqrt{9 a-B^2} \sqrt{6 a^2-B^4}\right)}{\sqrt{6 a^2-B^4}} \,, \label{phisol} \\
f(z) & = -\frac{2}{z^2} e^{2 B^2 z^2 + 2 A(z)} \sqrt{1 + \frac{B^2 z^4 e^{-2 B^2 z^2 - 4 A(z)} }{b^2 }} \nonumber \\
& \quad \times \left(-2 g(z)+ 2 B^2 z^2 g(z)+ 3 z  g(z) A'(z)+ z g'(z)\right)\,,  \label{fsol} \\
V(z) & = - b^2 \left(-1 + \sqrt{1 + \frac{B^2 e^{-2 B^2 z^2 - 4 A(z)} z^4}{b^2}} \right) f(z)+ \frac{1}{2} e^{-2A(z)}\bigl(-2 g(z) \left( 12 \right. \nonumber \\
& \quad -10 B^2 z^2  +4 B^4 z^4 \left.+ 6 z A'(z)\left(-3 +2 B^2 z^2 \right) +9 z^2 A'(z)^2 + 3 z^2 A''(z) \right) \nonumber \\
& \quad+ z \left(\left(9 - 6 B^2 z^2 - 9 z A'(z)\right)  g'(z) - z  g'(z) \right)\bigr) \ ,
\label{EBIsolutions}
\end{align}
where apart from the above-mentioned boundary conditions, we have further demanded that the dilaton field $\phi$ must be real-valued throughout the entire bulk and that it must go to zero at the asymptotic boundary $\phi(0)=0$ (asymptotically AdS). The above solution represents a black hole with a horizon at $z=z_h$ where the holographic radial coordinate is denoted by $z$, which varies from the asymptotic boundary ($z=0$) to the horizon radius ($z=z_h$). 

There exists another solution corresponding to thermal-AdS. The thermal-AdS solution has no horizon and can be obtained from the black hole solution by taking the limit $z_h\rightarrow\infty$. The thermal-AdS and black hole phases correspond to the confined and deconfined phases in the dual field theory, respectively \cite{Witten:1998zw}. Interestingly, as we will see later on, there can be a Hawking/Page type phase transition between the thermal-AdS and black hole solutions. This phase transition in the dual boundary field theory corresponds to the confined/deconfined phase transition. Therefore, by analyzing the effects of $B$ on the thermal-AdS/black hole phase transition temperature, we can obtain information about how the background magnetic field affects the deconfinement transition temperature holographically.
 
Let us also note the expressions of temperature and entropy associated with the black hole solution as they will be useful to discuss the deconfinement transition temperature. These are 
 \begin{eqnarray}
 	 T &=& \frac{z_{h}^{3} e^{-3A(z_h)-B^2 z_{h}^{2}}}{4 \pi \int_0^{z_h} \, d\xi \ \xi^3 e^{-B^2 \xi^2 -3A(\xi) } } \,,   \qquad S_{BH} ~=~ \frac{V_3 e^{3 A(z_h)+B^2 z_{h}^{2}}}{4 G_5 z_{h}^3 } \,,
 	\label{BHtemp}
 \end{eqnarray}
where $V_3$ denotes the volume of the three-dimensional spatial plane.

Notice that $a$ is the only arbitrary parameter left in the gravity solution in Eqs.~(\ref{asol})-(\ref{EBIsolutions}) and so the solution is completely determined once $a$ is fixed. Its value can be fixed by taking inputs from the dual boundary QCD theory. Following \cite{Dudal:2017max}, here we fixed its value by demanding the Hawking/Page (or the dual confinement/deconfinement) transition temperature to be around $270~\text{MeV}$ at zero magnetic field in the pure glue sector \cite{Lucini:2003zr}. This fixes $a$ to $0.15~\text{GeV}^{2}$. Also, notice from Eq.~(\ref{phisol}) that for a fixed $a$ the realness condition of the dilaton field places a constraint on the allowed values of $B$. For instance, for $a=0.15~\text{GeV}^2$, the largest allowed value of $B$ is $B\backsimeq 0.6~\text{GeV}$ \cite{Bohra:2020qom}. In the rest of the paper, we will take $a=0.15~\text{GeV}^{2}$ and $B \leq 0.6~\text{GeV}$.

\section{Setting the Born-Infeld parameter value $b$}
\label{sec3}
Let us now determine a suitable value for the Born-Infeld parameter $b$. It is related to the string length parameter $\alpha'$ via $b=1/(2 \pi \alpha')$. The value of $b$ can thus be fixed by computing the Wilson loop (string tension) holographically in our model and matching those with lattice QCD results. The Wilson loop is a gauge-invariant observable defined as the path-ordered exponential of the line integral of a gauge field along a closed loop $C$. The Wilson loop was unfortunately inadequate to fix the parameter $b$ in the earlier EBID-based soft-wall models \cite{Dudal:2014jfa}, as these models do not render the (expected) linear confinement potential between the quark-antiquark pair \cite{Karch2010Dec}. A great advantage of our self-consistent EBID model is that the Wilson loop now respects the area law in the confined phase, and therefore can be used to fix $b$. 

From the world sheet on-shell Nambu-Goto action, we can calculate the free energy of the quark-antiquark pair holographically. The free energy is given by,
 \begin{eqnarray}
 	\mathcal{F}(T,\ell)=T \ S_{NG}^{on-shell} \,.
 	\label{Sonshell}
 \end{eqnarray}
Here, $S_{NG}^{on-shell}$ is the on-shell Nambu-Goto action,
\begin{eqnarray}
	S_{NG}=\frac{1}{2 \pi \ell_{s}^2}\int d\tau d\sigma \sqrt{-\det g_s}, \ \ \ (g_s)_{\alpha\beta}=(g_s)_{\mu \nu}\partial_\alpha X^{\mu} \partial_\beta  X^{\nu} \,,
	\label{NGaction}
\end{eqnarray}
where $T_s=1/2 \pi \ell_{s}^2$ is the open string tension, $(\tau,\sigma)$ are the world sheet coordinates, $X^{\mu}(\tau,\sigma)$ represents the open string embedding, and $g_s$ denotes the background metric in the string frame. It is related to the Einstein frame metric in the following way,
\begin{eqnarray}
	& & (g_s)_{\mu \nu}=e^{\sqrt{\frac{2}{3}}\phi} g_{\mu \nu} \,,  \nonumber \\
	& & ds_{s}^2= \frac{L^2 e^{2 A_{s}(z)}}{z^2}\biggl[-g(z)dt^2 + \frac{dz^2}{g(z)} + dx_{1}^2+ e^{B^2 z^2} \left(dx_{2}^2+dx_{3}^2\right) \biggr] \,,
	\label{stringmetric}
\end{eqnarray}
where $A_{s}(z)=A(z)+\sqrt{\frac{1}{6}} \phi(z)$.

Then, the expectation value of the Wilson loop (or the string tension) is related to the quark-antiquark free energy and is given by
\begin{equation}
\braket{W(C)} \sim e^{S_{NG}^{on-shell}}  \,.
\end{equation}
From the Wilson loop, we can also compute the string tension at $B=0$ and compare it with the available lattice results. More precisely, we can use the estimate $\sqrt{\sigma} =0.43\pm0.02~\text{GeV}$, see the review \cite[below eq.(4.62)]{Bali:2000gf}. This is based on a quenched QCD simulation, compared to a Cornell potential which contains a confining linear piece $\sigma r$, similar to our holographic setup, see also \cite{Bohra:2019ebj}.

The central value $\sqrt{\sigma} =0.43~\text{GeV}$ corresponds to $b=0.044~\text{GeV}^2$, and to be conservative, we will consider a larger, namely $10\%$, variation around it to check the stability of our result. For completeness, the foregoing $b$ corresponds to $\ell_s/L=1.92$.

It is important to note that the Wilson loop is usually calculated in scenarios involving heavy quark pairs. This approach helps circumvent the complexities associated with dynamical string breaking. Therefore, it is logical to establish the new scale, $\alpha'$, using the action related to heavy charm quarks within the EBID framework. Comparing our numerical results with lattice QCD results gives us a small value of $L/\ell_s$, suggesting a small 't Hooft coupling \cite{Mahapatra:2019uql}. This small value suggests that the classical gravity approximation may not fully hold and hints at the potential importance of higher derivative $\alpha'$ correction terms. While this discrepancy is a limitation of our model, it is worth noting that similar issues exist in most bottom-up gauge/gravity models \cite{Noronha2009Oct, Gubser2008Apr, Andreev2006Apr}, see also \cite{Rougemont:2023gfz} for a recent discussion. Models showing higher values of the ratio $L/\ell_s$ often have additional complexities, such as scaling symmetries in the dilaton potential or extra parameters, which raise questions on the validity of such high values \cite{Gursoy2009Mar}. Resolving this discrepancy requires further investigation. Indeed, the precise value of this ratio depends on selecting specific loop-related QCD observables. While our model's quantitative predictions may not match exactly with experimental values, we believe our qualitative conclusions remain valid and reliable. In particular, most of our holographic QCD predictions remain robust for different values of parameters $b$ and $a$, we numerically verified that the magnitudes of observables under consideration change only by a few tens of MeV, within the range permitted by lattice QCD. In the rest of the paper, without losing any generality, we will therefore mostly take $b=0.044~\text{GeV}^2$, that is, the value that reproduces best the string tension at vanishing temperature and magnetic field. 

\section{Black hole thermodynamics: a short survey}
\label{sec4}
In the considered magnetized EBID model, two distinct gravity solutions emerge. The first solution is a black hole solution featuring a horizon at $z=z_h$. The second solution corresponds to the thermal-AdS. Due to the existence of multiple solutions, we will analyze their global thermodynamic stabilities. Since both computations and results are rather analogous to the earlier studied Einstein-Maxwell-dilaton system, see for instance \cite{Dudal:2017max,Bohra:2019ebj,Jena:2022nzw}, we will be deliberately brief here.

\begin{figure}[htb!]
\begin{minipage}[b]{0.5\linewidth}
\centering
\includegraphics[width=2.8in,height=2.3in]{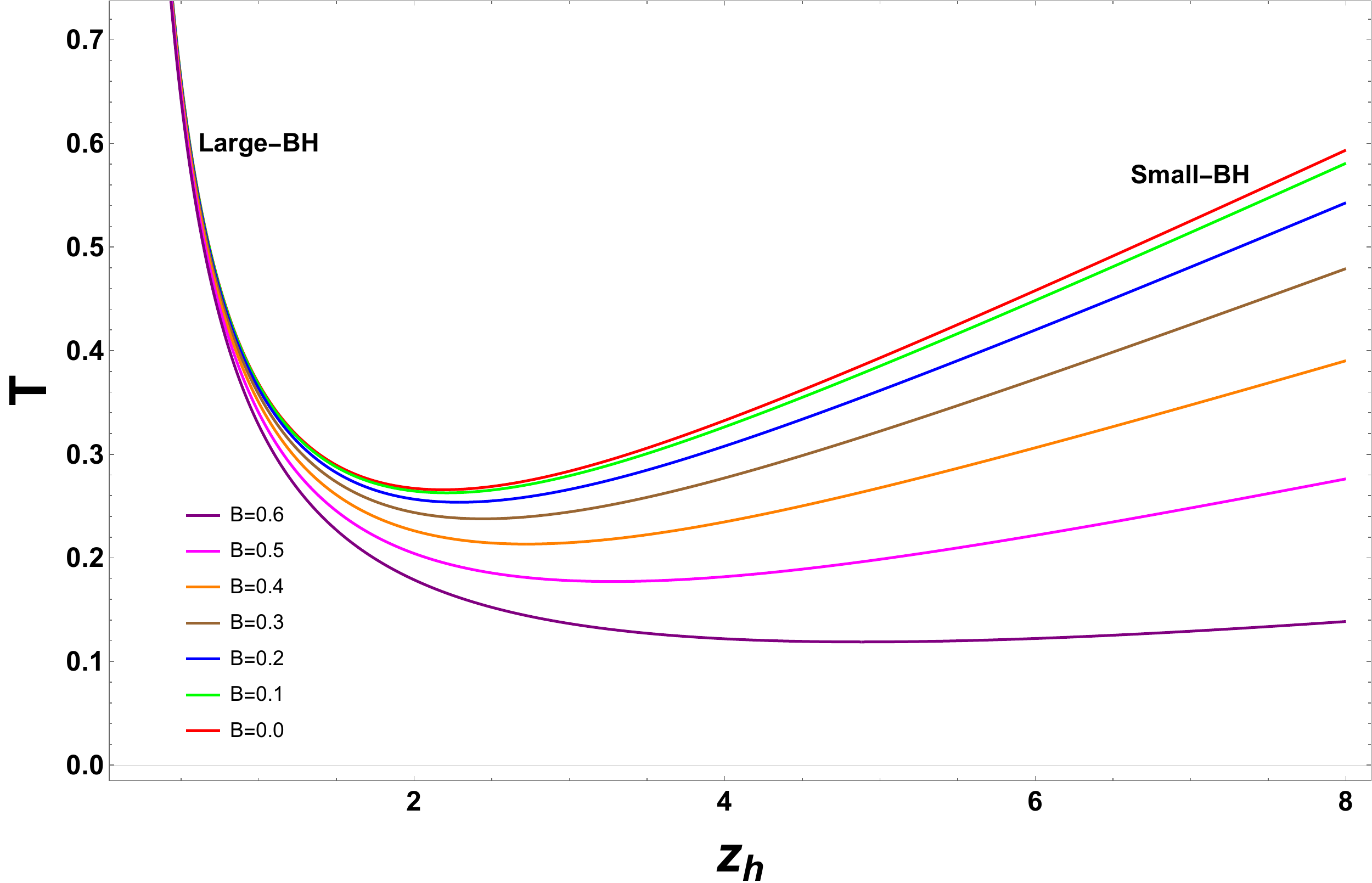}
\caption{\small Temperature $T$ as a function of horizon radius $z_h$ for various values of the magnetic field $B$. In units GeV.}
\label{TvszhBH}
\end{minipage}
\hspace{0.4cm}
\begin{minipage}[b]{0.5\linewidth}
\centering
\includegraphics[width=2.8in,height=2.3in]{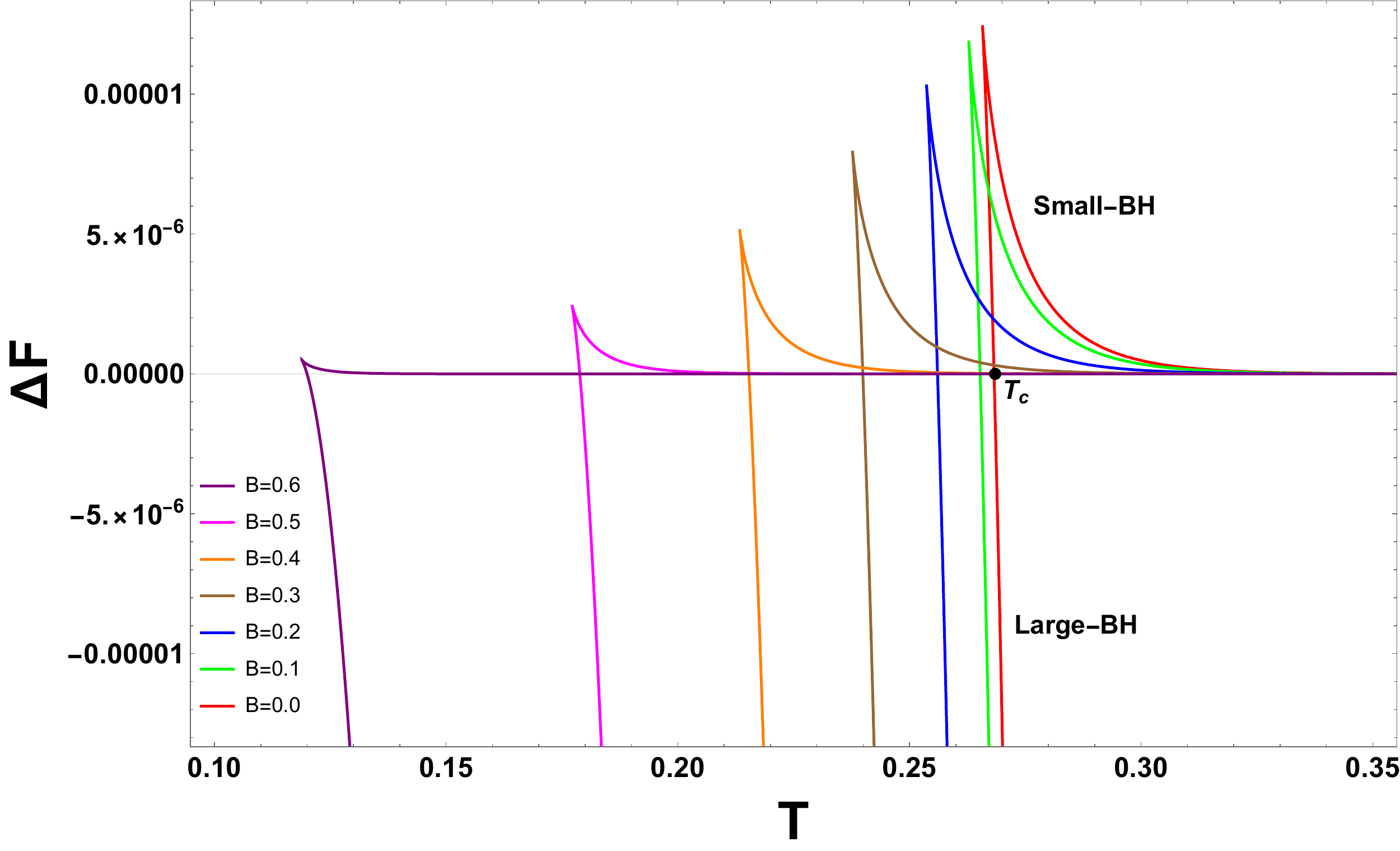}
\caption{\small The free energy difference $\Delta F$ as a function of temperature $T$ for various values of the magnetic field $B$. In units GeV.}
\label{FvsTBH}
\end{minipage}
\end{figure}
 
In Fig.~\ref{TvszhBH}, we illustrate the dependence of the Hawking temperature on the horizon radius for different magnetic field strengths. While the thermal AdS phase exists at every temperature, we observe the presence of a minimum temperature below which no black hole solution exists. Above the identified minimum temperature, the system exhibits two distinct black hole solutions: a large and a small black hole. The large black hole solution (small $z_h$) is characterized by a decreasing temperature trend with respect to the horizon radius. It has a positive specific heat and is therefore thermodynamically stable. In contrast, the small black hole solution (large $z_h$) displays an increasing temperature trend with horizon radius. It has a negative specific heat, rendering it thermodynamically unstable.

In Fig.~\ref{FvsTBH}, the free energy difference $\Delta F$ between the black hole and thermal-AdS phases is shown. Clearly, the free energy of the large black hole and thermal-AdS phases exchange dominance as the temperature is varied. In particular, the free energy of the large black hole phase is lowest at high temperatures, suggesting it to be a true global minimum in the high-temperature regime, whereas the free energy of the thermal-AdS phase is lowest at low temperatures, suggesting it to be a global minimum at low temperatures. On the other hand, the free energy of the small black hole phase is always higher than the thermal-AdS as well as the large black hole phases, making it thermodynamically disfavored at all temperatures. Accordingly, there is a first-order phase transition between the large black hole and thermal-AdS phases as the temperature is varied. This is nothing else than  the Hawking/Page phase transition \cite{Hawking1983Dec}.  The temperature at which the free energy difference goes to zero gives the corresponding transition temperature $T_c$. At $B=0$, the transition temperature is $270~\text{MeV}$, as determined earlier.

\begin{figure}[htb!]
	\centering
	\includegraphics[height=7cm,width=11cm]{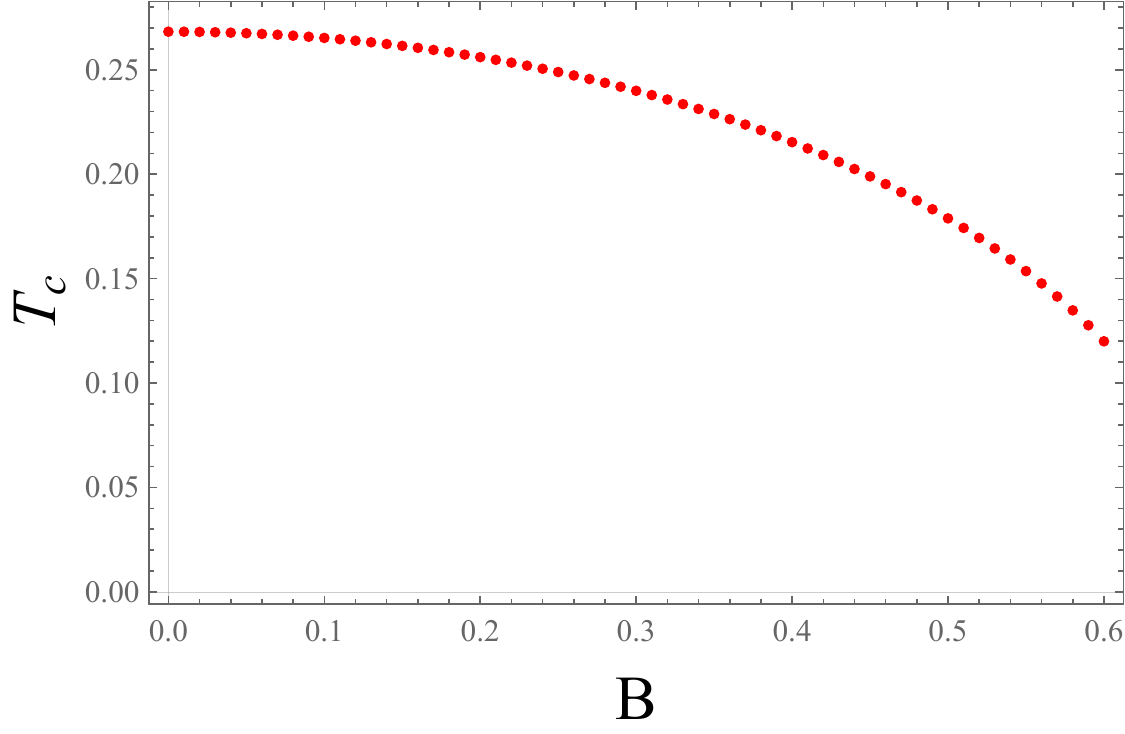}
	\caption{\small The variation of the deconfined transition temperature $T_c$ with magnetic field. In units of GeV.}
	\label{TcvsB}	
\end{figure}

Intriguingly, the described thermodynamic characteristics persist even for small yet finite magnetic field values ($B\leq 0.6$). The EBID gravity system continues to exhibit unstable small and stable large black hole phases, with the thermal-AdS phase dominating at low temperatures. Notably, the Hawking-Page thermal-AdS/black hole phase transition is still present at finite magnetic fields. The notable difference lies in the transition temperature, which decreases monotonically with higher magnetic field values. The dependence of $T_c$ on $B$ is illustrated in Fig.~\ref{TcvsB}. 

As mentioned earlier, the thermal-AdS and black hole phases have been established as dual counterparts to confined and deconfined phases, respectively, in the boundary theory. The above thermodynamic results therefore predict inverse magnetic catalysis in the deconfinement sector. This holographic result is in line with the lattice results \cite{Bali:2011qj}. The constructed EBID gravity also predicts, though the explicit results are not presented here for brevity,  an decrement/increment in the quark-antiquark string tension with the magnetic field when the pair is aligned parallel/perpendicular to it, thereby once again correlating well with recent lattice results \cite{Bonati:2014ksa,Bonati:2016kxj}. It is important to emphasize that these holographic QCD features in the presence of $B$ are quite robust and remain qualitatively the same for different values of $a$ and $b$, albeit with different magnitudes of $T_c(B)$.

As our main focus is on investigating the magnetic field effects on the melting of quarkonium, a process that involves the dissociation of quark-antiquark bound states and is intricately linked to the deconfinement sector, we now advance our calculations by considering the deconfined black hole phase for further analysis.

\section{Equations of motion for longitudinal and perpendicular quark\-onium fluctuations}
\label{sec5}
To study the spectral functions, we need to first write down the equations of motion of the relevant gauge field fluctuations. These can be obtained from the Born-Infeld Lagrangian. From the EBID action (\ref{actionEBI}), the Born-Infeld (BI) part of the Lagrangian is,
\begin{equation}
\mathcal{L}_{BI} = b^2 \sqrt{-g} f(\phi) \left[1-\sqrt{1+ \frac{F_{\mu \nu} F^{\mu \nu}}{2 b^2}}\right].
\label{BIlagrangian}
\end{equation}
This can also be rewritten as (see Appendix A),
\begin{equation}
	\mathcal{L}_{BI} = b^2 f(\phi)\left(\sqrt{-g}-\sqrt{-\det\left(g_{\mu \nu}+ \frac{F_{\mu \nu}}{b}\right)}\right)\,.
	\label{BIlagrangiansecond}
\end{equation}
Now we decompose the gauge field into a background term plus a fluctuation term, i.e., $F_{\mu \nu}= \bar{F}_{\mu \nu} + \tilde{F}_{\mu \nu}$. Then the second term of Eq.~(\ref{BIlagrangiansecond}) can be expanded as
\begin{eqnarray}
\sqrt{-\det\left(g_{\mu \nu}+ \frac{F_{\mu \nu}}{b}\right)} = \sqrt{-\det\left(g_{\mu \nu} + \frac{1}{b} (\bar{F}_{\mu \nu} + \tilde{F}_{\mu \nu})\right)}  = \sqrt{-\det\left(\mathcal{G}_{\mu \nu}+\frac{\tilde{F}_{\mu \nu}}{b}\right)}\,, \nonumber\\
\approx \sqrt{-\det(\mathcal{G})}\left\{1+\frac{1}{2b}\text{Tr}(\mathcal{G}^{-1}\tilde{F}) + \frac{1}{8 b^2}(\text{Tr}(\mathcal{G}^{-1}\tilde{F}))^2-\frac{1}{4 b^2} \text{Tr}((\mathcal{G}^{-1}\tilde{F})^2)+....\right\}\,,
\label{BIlagrangianexpansion}
\end{eqnarray}
where $\mathcal{G}_{\mu \nu}=g_{\mu \nu} + \bar{F}_{\mu \nu}/b$. Explicitly, recalling  the background gauge field Ans\"{a}tze~(\ref{ansatze}),
\begin{equation}
	\bar{F}_{\mu \nu}=B dx_{2}\wedge dx_{3}\,,
	\label{fieldtensoransatze}
\end{equation}
the metric tensor $\mathcal{G}_{\mu \nu}$ is given by,
\begin{equation}
\mathcal{G}_{\mu \nu} = \begin{bmatrix}
	g_{00} & 0 & 0 & 0 & 0\\
	0 & g_{zz} & 0 & 0 & 0\\
	0 & 0 & g_{11} & 0 & 0\\
	0 & 0 & 0 & g_{22} & \frac{B}{b}\\
	0 & 0 & 0 & -\frac{B}{b} & g_{g_{33}} 
\end{bmatrix} \,.
\label{metrictensor}
\end{equation}
It has a determinant 
\begin{equation}
	\mathcal{G} = g_{00} g_{11} g_{zz} \left(g_{22} g_{33} + \frac{B^2}{b^2}\right)\,,
	\label{metricdeterminant}
\end{equation}
and its inverse is given by
\begin{equation}
	\mathcal{G}^{\mu \nu} = \begin{bmatrix}
		\frac{1}{g_{00}} & 0 & 0 & 0 & 0\\
		0 & \frac{1}{g_{zz}} & 0 & 0 & 0\\
		0 & 0 & \frac{1}{g_{11}} &0 & 0\\
		0 & 0 & 0 &\frac{g_{33}}{X} &  -\frac{B}{b X}\\
		0 & 0 & 0 & \frac{B}{b X} & \frac{g_{22}}{X}
	\end{bmatrix}\,,
\label{metricinverse}
\end{equation}
where $X=g_{22} g_{33} + \frac{B^2}{b^2}$. It is convenient if we decompose the metric tensor $\mathcal{G}_{\mu \nu}$ into the symmetric ($G$) and antisymmetric ($S$) parts,
\begin{equation}
\begin{aligned}  
    G^{\mu \nu} = \begin{bmatrix}
        \frac{1}{g_{00}} & 0 & 0 & 0 & 0\\
        0 & \frac{1}{g_{zz}} & 0 & 0 & 0\\
        0 & 0 & \frac{1}{g_{11}} & 0 & 0\\
        0 & 0 & 0 &\frac{g_{33}}{X} & 0\\
        0 & 0 & 0 & 0 & \frac{g_{22}}{X}
    \end{bmatrix}\,,
    \qquad
    S^{\mu \nu} = \begin{bmatrix}
        0 & 0 & 0 & 0 & 0\\
        0 & 0 & 0 & 0 & 0\\
        0 & 0 & 0 & 0 & 0\\
        0 & 0 & 0 &0 & -\frac{B}{b X}\\
        0 & 0 & 0 & \frac{B}{b X} & 0
    \end{bmatrix}\,.
    \label{symantisym}
\end{aligned}
\end{equation}
To study the spectral function we concentrate on vector modes $V = \frac{A_L + A_R}{2}$ and select the gauge $A_z = 0$ \cite{Dudal:2015kza}. Subsequently, we employ a Fourier expansion of the modes $\sim e^{i\textbf{q} \cdot \textbf{x} -i \omega t}$. However, to simplify the numerics, in the rest of the paper, we will concentrate on the case when there is no momentum on the boundary, i.e., $\textbf{q}=0$.\par
The fluctuation equations of motion can be obtained from the gauge field equation (\ref{maxwelleom})
\begin{eqnarray}
\partial_{\mu}\biggl[\sqrt{-g}\frac{f(\phi)}{\sqrt{1+\frac{(\bar{F}_{\mu \nu} + \tilde{F}_{\mu \nu})(\bar{F}^{\mu \nu} + \tilde{F}^{\mu \nu})}{2b^2}}} (\bar{F}^{\mu \nu} + \tilde{F}^{\mu \nu})\biggr]=0\,,
\label{fluctuationeom}
\end{eqnarray}
where we have substituted  $F_{\mu \nu}= \bar{F}_{\mu \nu} + \tilde{F}_{\mu \nu}$. At leading (linear) order, the equation of motion for the fluctuation is given by
\begin{equation}
	\partial_z^2 V_i + \partial_z \left(\ln\left(\sqrt{-\mathcal{G}} f(z) G^{zz} G^{ii}\right)\right)\partial_z V_i - \frac{G^{tt}}{G^{zz}} \omega^2 V_i = 0\,,
 \label{parallelfluceq}
\end{equation}
where $i=1$ corresponds to the fluctuation in the parallel direction, whereas $i=2$ (or $3$) corresponds to the fluctuation in the perpendicular direction. 
Interestingly, as in \cite{Dudal:2014jfa}, one can obtain the same equation of motion for $V_1$ and $V_2$ from the following modified gauge equation
\begin{equation}
\partial_{\mu}\left(f(\phi)\sqrt{-\mathcal{G}} \tilde{F}^{\mu \nu}\right)=0\,.
\label{fluctuationEOM}
\end{equation}
With these expressions in hand, it is straightforward to see that the term $\text{Tr}(\mathcal{G}^{-1}\tilde{F})$ in Eq.~(\ref{BIlagrangianexpansion}) vanishes by the equation of motion of the background gauge field. The remaining terms exhibit quadratic behavior in the fluctuations. We can further decompose the inverse background $\mathcal{G}^{-1}$ into its symmetric and antisymmetric parts. The symmetric part is solely influenced by the final term, constituting the Maxwell action where indices are raised and lowered using just the symmetric component. In this way, one can show that (\ref{fluctuationeom}) and (\ref{fluctuationEOM}) are the same as far as the equations of motion of the fluctuations are considered.

\section{Computation of the spectral function based on the real-time AdS/CFT dictionary}
\label{sec6}
\subsection{Asymptotic behavior: Frobenius analysis and power expansion}
In order to compute the spectral functions numerically we adhere to the prescription that is shown in \cite{Miranda:2009uw}, which is itself based on the seminal work \cite{Son:2002sd} that describes how to compute, in a AdS/CFT setting, correlators in Minkowski spacetime. The first step of the procedure is to analyze the asymptotic structure of the solutions $V_i$. To do so we compute the horizon and near boundary expansion of $V_i$, which both can be obtained from the Frobenius analysis and a power expansion, respectively.

For the case where the fluctuation is parallel to the background magnetic field, the relevant equation of motion is given by
\begin{eqnarray}
	\partial_z^2 V_1 + \partial_z \left(\ln\left(\frac{e^{-A(z)} f(z) g(z)}{L z} \sqrt{e^{2 B^2 z^2+ 4 A(z)} L^4 + 4 B^2 \pi^2 z^4 \alpha{'}^2} \right)\right)\partial_z V_1 + \frac{\omega^2}{g(z)^2}  V_1 = 0\,.
\end{eqnarray}
If, however, the fluctuation is perpendicular to the magnetic field, we have the expression
\begin{eqnarray}
	&&\partial_z^2 V_2 + \partial_z \left(\ln\left(\frac{L^3 f(z) g(z) e^{3 A(z)+B^2 z^2} \sqrt{L^4 e^{4 A(z)+2 B^2 z^2}+4 \pi ^2 \alpha{'} ^2 B^2 z^4}}{L^4 z e^{4
   A(z)+2 B^2 z^2}+B^2 z}  \right)\right)\partial_z V_2 \nonumber\\&&+ \frac{\omega^2}{g(z)^2}  V_2 = 0 \,.
\end{eqnarray}

All the analytical and numerical analysis that will follow is done for these two equations of motion.

\subsubsection{Near horizon regime: power expansion}

As shown in other works \cite{Miranda:2009uw, Mamani:2018uxf}, it is convenient to switch to a Regge-Wheeler tortoise coordinate $r_*$ to investigate classical perturbations in a background black hole spacetime. In general, the tortoise coordinate is defined by $\partial_{r_*} = -g(z) \partial_z$. Ordinarily, in a non-dynamical soft-wall model, in which Einstein's equations are not satisfied, $1/g(z)$ can be explicitly integrated and one can find a closed expression for $r_*$. As an example, for a simple $g(z) = 1 - \frac{z^4}{z_h^4}$, one finds
\begin{eqnarray}
	r_*(z) = \frac{z_h}{2} \left[\log\left({\frac{z_h-z}{z_h+z}}\right) -\arctan\left({\frac{z}{z_h}}\right) \right]\,,
\end{eqnarray}
However, our expression for $g(z)$ is quite complicated and we did not manage to find a closed expression for $r_*$. Consequently, in order to proceed with the computations, we do a series expansion of $1/g(z)$ up to a given order and integrate the given result. Up to the second order this procedure gives an approximate $r_*$ given by
\begin{dmath}
  r_*(z) \approx  \frac{z_h^2 \left(3 a-B^2\right)+e^{z_h^2 \left(B^2-3 a\right)}-1}{144 z_h^6 \left(B^2-3a\right)^2} \left[z \left(-4 z_h^4 \left(3 a-B^2\right) \left(z^2 \left(3 a-B^2\right)-30\right)+24 z z_h^3
   \left(B^2-3 a\right)+12 z_h^6 \left(B^2-3 a\right)^2+15 z^2-90 z z_h+243 z_h^2\right)-144
   z_h^3 \tanh ^{-1}\left(\frac{z}{z-2 z_h}\right) \right]\,.
\end{dmath}
Notice that $g(z_h) = 0$ and hence we see that the equations of motion take a free particle form near the horizon. Therefore, it is possible to express the solutions, close to the horizon, as ingoing and outgoing plane waves traveling in and out of the black hole interior, 
\begin{eqnarray}
	\psi \sim \mathcal{C}e^{-i\omega r_*} + \mathcal{D}e^{i\omega r_*} \,,
\end{eqnarray}
Knowing this, the solutions to the equations of motion near the horizon can be expressed by the following power expansion
\begin{eqnarray}
	\psi_{\pm} = e^{\pm i\omega r_*} \left[a_0^{\pm} + a_1^{\pm} (z_h - z) + a_2^{\pm} (z_h - z)^2 + \dots \right] \,.
\end{eqnarray}
where we choose $a_0^{\pm} = 1$. After we substitute the above expression into the equations of motion, for the parallel and perpendicular cases, we can solve for the $a_1^{\pm}$ and $a_2^{\pm}$ coefficients. Given the complexity of the equations of motion and the presented tortoise coordinate, these coefficients are quite complicated, and we choose to not show them here. Suffice to say that they are functions of $\omega$ and the parameters of the model: $B$, $a$, etc. For further computations, the solutions $\psi_{\pm}$ are truncated at order $(z_h - z)^2 $.

Now, clearly, the in-~and outgoing solutions form a basis for any other wave function we might consider. Therefore, renormalizable and non-renormalizable solutions, respectively, can be written as a linear combination of the in-~and outgoing solutions:
\begin{align}    
	\psi_{2} &= \mathcal{C}_2 \psi_- + \mathcal{D}_2 \psi_+ \,, \\
	\psi_{1} &= \mathcal{C}_1 \psi_- + \mathcal{D}_1 \psi_+ \,.
\end{align}

\subsubsection{Near boundary regime: Frobenius analysis}
In the limit $z \rightarrow 0$, we take the first term in the power series solution as $z^\beta$, as the prescription for the Frobenius expansion (near $z=0$). The asymptotic analysis of Eqs.~(\ref{parallelfluceq}) then simply gives the following indicial equation
\begin{equation}
	\beta(\beta-4)=0\,,
\end{equation}
yielding $\beta=0$ or $\beta=4$. Since we would like to solve the fluctuation differential equations by evolving from the boundary to the horizon, we also need the information on the derivative of $V_i$ at the boundary to consistently set up the numerical routine. Obtaining the first derivative of the solution at the holographic boundary requires a deeper exploration of the asymptotic structure of $V_i$ by Frobenius analysis. Since the indicial equation yields two solutions with an integer difference, the larger solution ($\beta=4$) can have a regular series expansion, while the other solution manifests as a logarithmic series. These two solutions are represented by
\begin{align}
    \psi_1(z)  &= z^4 \sum_{k=0}^{+\infty} a_k z^k \,, \\
    \psi_2(z) &= C \ln(z) \psi_1(z)+ \sum_{k=0}^{+\infty} b_k z^k \,.
 \label{power}
\end{align}
The explicit forms of $a_k$ and $b_k$ can be found by substituting the above series solution into Eqs.~(\ref{parallelfluceq}) and equating the terms order by order. As normalization, we choose $a_0=b_0=1$. The coefficients of the expansions are functions of $\omega$ and the parameters of the model: $B$, $a$, etc. As an example, we find that the first coefficient, for the considered equations of motion, of $\psi_1(z)$ can be written as
\begin{align}
        a_1^{\text{parallel}} &= \frac{1}{12} \left(-16 B^2-\omega ^2\right) \,, \\
        a_1^{\text{perpendicular}} &= \frac{1}{12} \left(96 a^2+96 a B^2-24 a-8 B^2-\omega ^2\right)	\,.
\end{align}  
These solutions starting from near the boundary also provide a basis for constructing wave functions,
\begin{align}
	\psi_{+} &= \mathcal{A}_+ \psi_2 + \mathcal{B}_+ \psi_1 \,, \\
	\psi_{-} &= \mathcal{A}_- \psi_2 + \mathcal{B}_- \psi_1 \,.
\end{align}

\subsection{Retarded Green's function}
Both near boundary and horizon expansions give a basis in which wave functions can be written as linear combinations. We can see it right away from the expressions for $\psi_{\pm}$, $\psi_{1}$ and $\psi_{2}$ that the coefficients follow the relation
\begin{gather}
 \begin{pmatrix} \mathcal{A}_{+} & \mathcal{B}_{+} \\ \mathcal{A}_{-} & \mathcal{B}_{-} \end{pmatrix}
 =
  \begin{pmatrix} \mathcal{C}_{2} & \mathcal{D}_{2} \\ \mathcal{C}_{1} & \mathcal{D}_{1}
   \end{pmatrix}^{-1} \,.
\end{gather}
We are interested in the computation of the spectral functions. To do such, in this approach, we need to find the retarded Green's function and take its imaginary part \cite{Mamani:2022qnf}. 

Following the procedure elucidated in \cite{Miranda:2009uw}, the computation of the retarded Green's function begins in numerically integrating the equations of motion, in a regular lattice that begins at a point near the boundary $z = \epsilon$ and ends at a point near the horizon $z = z_h - \epsilon$, with initial conditions given by the solution near the boundary, $z = 0$, $\psi_1(\epsilon)$, computed from the Frobenius Analysis, and its derivative $\partial_z \psi_1(\epsilon)$.

The other solution found near the boundary $\psi_2(\epsilon)$ and its derivative $\partial_z \psi_2(\epsilon)$ are also employed as initial conditions in the same numerical procedure described above to compute the other solution for the equations of motion.  

Given so, we are left with two numerical results for $\psi$. These functions are evaluated near the horizon $z_h - \epsilon$ in the next step, where we have the following construction:
\begin{gather}
 \begin{pmatrix} \psi_a(z_h - \epsilon) \\ \partial_z \psi_a(z_h - \epsilon) \end{pmatrix}
 =
  \begin{pmatrix} \psi_-(z_h - \epsilon) & \psi_+(z_h - \epsilon) \\ \partial_z  \psi_-(z_h - \epsilon) & \partial_z \psi_+(z_h - \epsilon)
   \end{pmatrix}
   \begin{pmatrix} \mathcal{C}_a \\ \mathcal{D}_a \end{pmatrix}\,,
\label{coef}
\end{gather}
where $a = 1,2 $ runs for the numerical results obtained in solving the equations of motion and $\psi_{\pm}$ are the asymptotic solutions near the horizon given by the power expansions in Eq.~(\ref{power}). Everything is evaluated at $z_h - \epsilon$. The coefficients $C_a$ and $D_a$ are then computed by inverting the above square matrix. If we combine this with the relation for the coefficients seen in Eq.~(\ref{coef}) we get 
\begin{eqnarray} \label{green}
    \frac{\mathcal{B}_-}{\mathcal{A}_-} = - \frac{\mathcal{D}_2}{\mathcal{D}_1} = \frac{\partial_z  \psi_-(z_h - \epsilon) \psi_2(z_h - \epsilon) - \psi_-(z_h - \epsilon)\partial_z \psi_2(z_h - \epsilon)}{\partial_z  \psi_-(z_h - \epsilon)\psi_1(z_h - \epsilon) -  \psi_-(z_h - \epsilon)\partial_z \psi_1(z_h - \epsilon)}\,.
\end{eqnarray}
This gives a stable numerical procedure to compute the ratio $\frac{\mathcal{B}_-}{\mathcal{A}_-}$ and, actually, this is all the information needed to compute the retarded Green's function and, consequentially, the spectral function. As it is shown in works \cite{Miranda:2009uw, Mamani:2018uxf, Mamani:2022qnf}, the spectral functions are directly related to $\frac{\mathcal{B}_-}{\mathcal{A}_-}$ by the expression
\begin{eqnarray}
    \rho(\omega,q=0) = - 2 \Im{G^R(\omega,q=0)} \propto \Im{\frac{\mathcal{B}_-}{\mathcal{A}_-}} \,.
\end{eqnarray}
where $ \rho(\omega,q=0)$ is the spectral function calculated numerically from the described prescription (where $q=0$ was set from the start), $\Im$~denotes the imaginary part.

The numerical integration of the equations of motion over the interval $z \in [\epsilon, z_h - \epsilon]$, to compute $\psi_i(z_h - \epsilon)$ and $\partial_z \psi_i(z_h - \epsilon)$ appearing in the formula for the Green's function (\ref{green}), is done by applying the ``StiffnessSwitching'' numerical scheme for NDSolve as provided by Mathematica. This was chosen as we recognized the problem to have a stiff nature (explicit methods like regular 4th order Runge-Kutta or Euler were unable to give sensible results). The other pieces of (\ref{green}), $\psi_{\pm}(z_h - \epsilon)$ and $\partial_z \psi_{\pm}(z_h - \epsilon)$, are actually computed from the analytical results of the Frobenius analysis and power expansions.

The small positive number $\epsilon$ was also used as the lattice spacing and the value chosen was $\epsilon = 0.001$ for the actual calculations. $z_h$, for a given temperature $T$, is computed by finding the root, using Newton's method (as the equation is transcendental in $z_h$), of the equation
\begin{eqnarray}
    T = - \frac{g'(z_h)}{4 \pi}\,,
\end{eqnarray}
for the horizon function $g(z)$ presented in Eq.~(\ref{gsol}). 

\subsection{Numerical results}
The spectral function describes the distribution of states that can be excited by a quarkonium state as it dissolves in a hot and dense QCD medium. It also captures information about the dissociation of the quarkonium states and QGP formation. In particular, the spectral peaks represent the presence of bound states of heavy quarks and anti-quarks (such as charm and bottom quarks) within the QGP. The presence, absence, or modification of these peaks can provide valuable information about the properties of the QGP. From above we see that by employing the small $\epsilon$ limit, we can effectively extract valuable information about the spectral function holographically.

The experimental measurements and theoretical observations of quarkonia production and suppression in heavy-ion collisions are thus not just an important tool for studying the properties of QGP but also for understanding the nature of the strong interactions under extreme conditions. Here, we are interested in observing how the quarkonia melts by looking at how the spectral peak changes when there is both temperature and background magnetic field.

\subsubsection{Parallel case}
\begin{figure}[htb!]
	\centering
 \subfigure[\pmb{$T=T_c$}]{\includegraphics[width=0.45\linewidth]{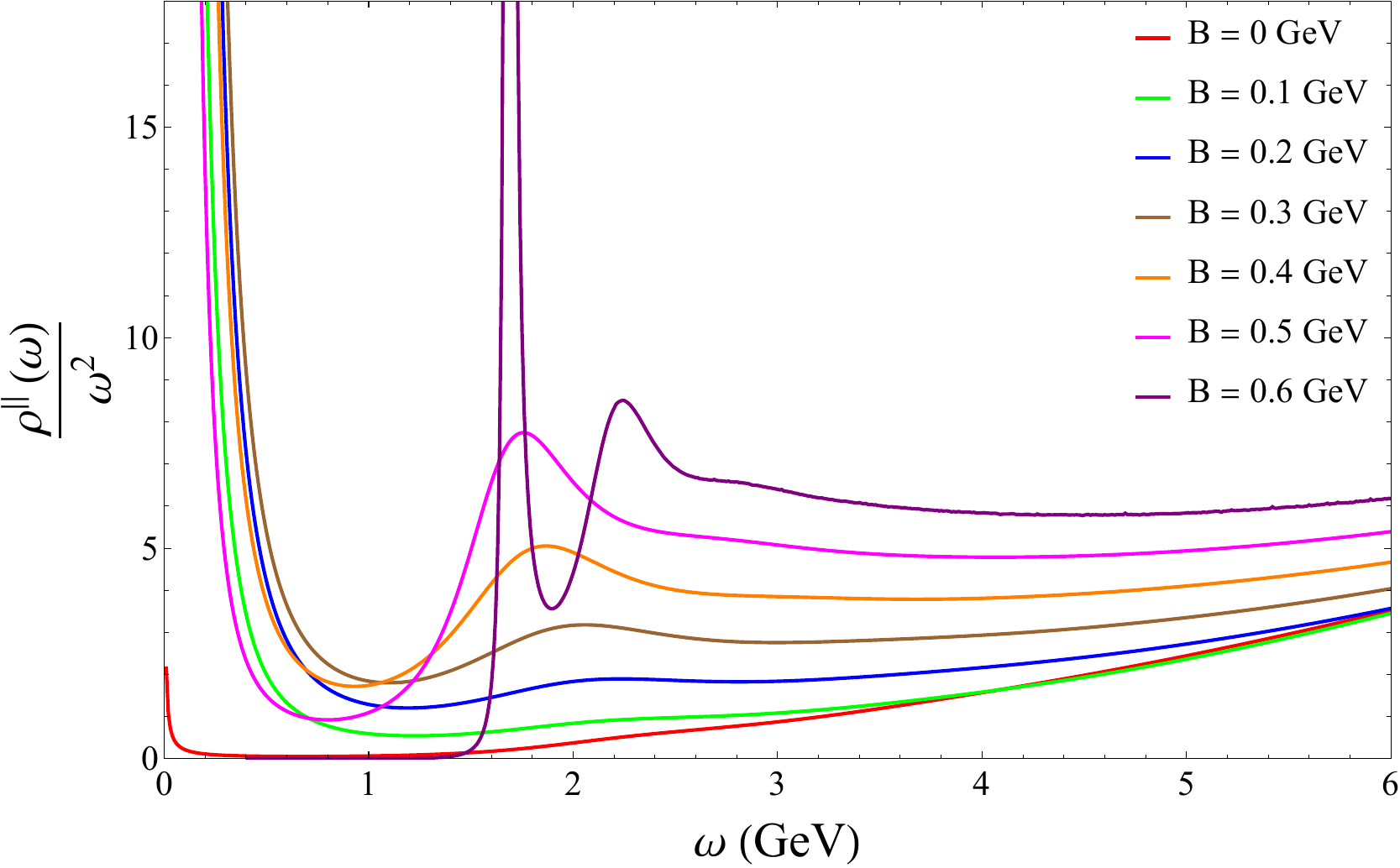}}
  \subfigure[\pmb{$T=1.5T_c$}]{\includegraphics[width=0.45\linewidth]{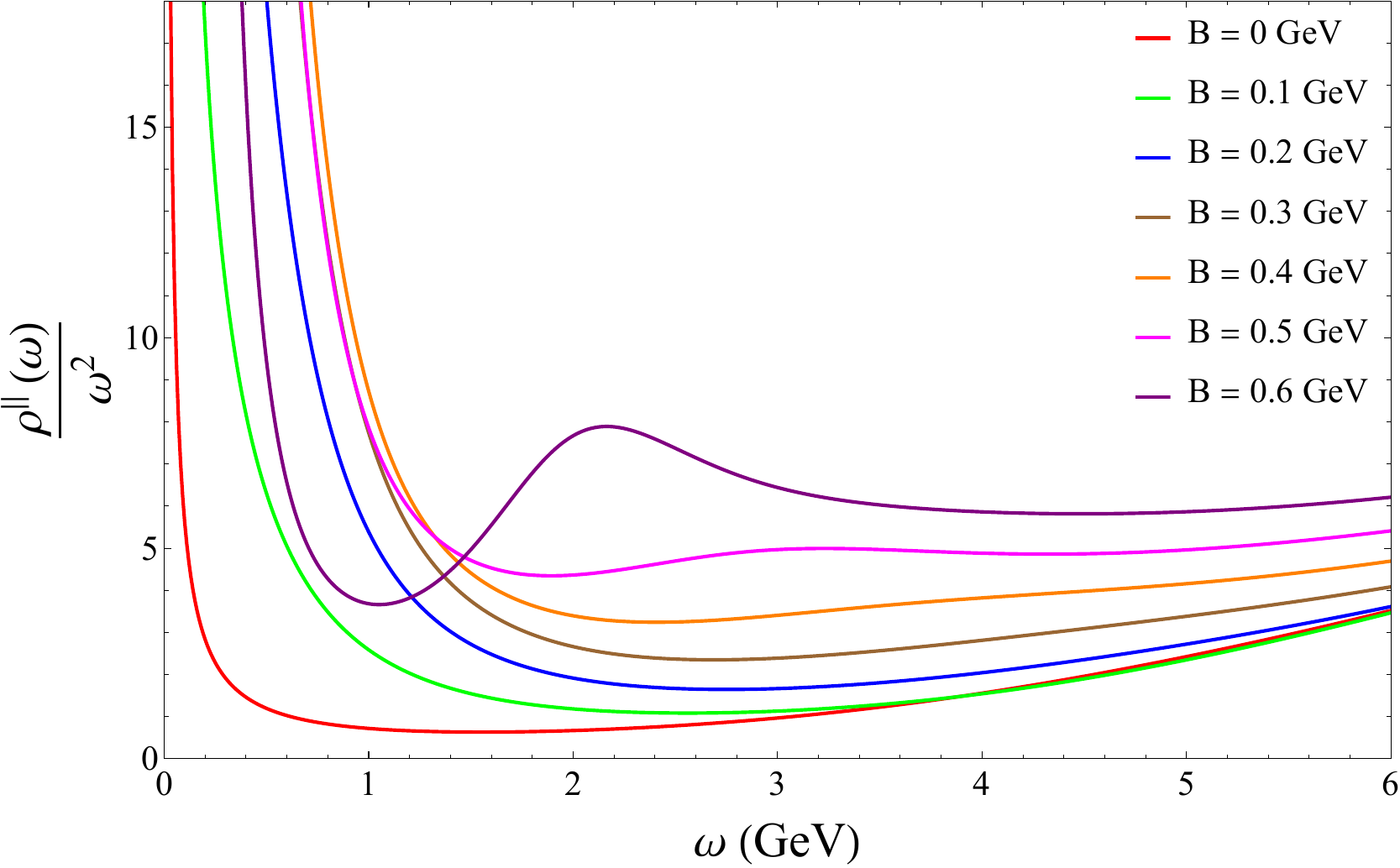}}
	\caption{\small Spectral function as a function of $\omega$ in the parallel direction for different magnetic fields. The temperature is fixed at $T=T_c$ (left) and $T=1.5~T_c$ (right). }
	\label{spf_diffB_parallel}	
\end{figure}
First, we focus on the scenario where the polarization aligns with the background magnetic field. The parametrized spectral function as a function of $\omega$ is illustrated in Fig.~\ref{spf_diffB_parallel} for different background magnetic field strengths at the deconfinement temperature $T=T_c$ and at $T=1.5~T_c$.\footnote{In Fig.~\ref{spf_diffB_parallel}, $\rho/\omega^2$ (instead of $\rho$) is plotted to make comparisons with lattice QCD results \cite{Ding2019Feb,Petreczky2006Jan}. Note that these lattice results have not considered the background magnetic field's impact.} Note the broadening of the spectral peaks at lower $B$ values, while, conversely, at higher magnetic fields, there is an emergence of multiple spectral peaks, indicating less influence of the magnetic field on the quarkonium states, which is similar to the results of \cite{Zhao2023Jun}. This in turn indicates that the magnetic field hinders the dissociation of the bound state, a scenario suggesting magnetic catalysis. Interestingly, lattice simulations as in \cite{DElia:2021tfb} also suggest an amplified string tension relative to the magnetic field's orientation, so one might expect to see traces of this anisotropic confinement reflected in the quarkonium spectra.

\begin{figure}[htb!]
  \centering
  \subfigure[$\pmb{B=0.0}$]{\includegraphics[width=0.45\linewidth]{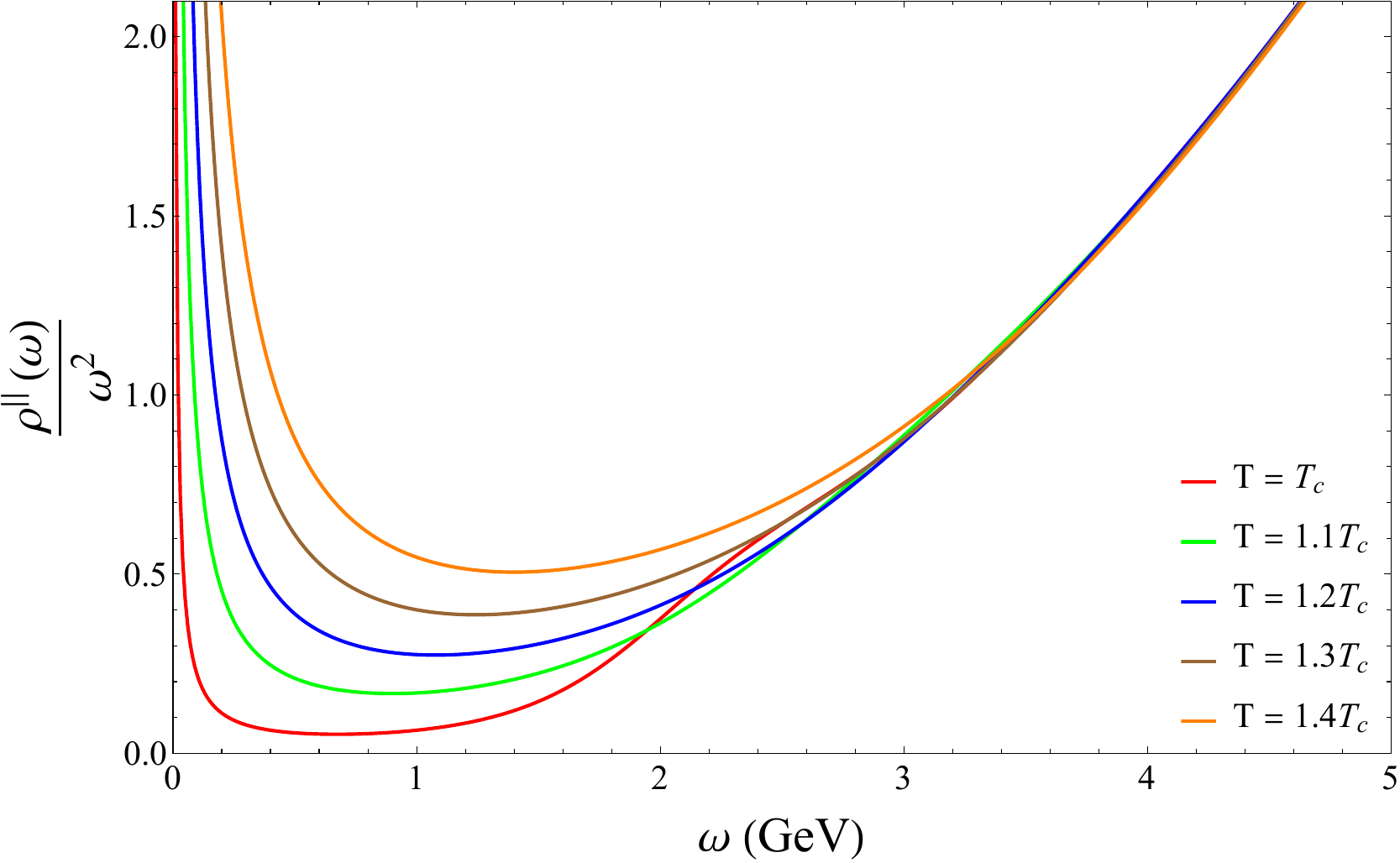}}
  \label{spf_fixedB_parallelB0}
  \subfigure[$\pmb{B=0.2}$]{\includegraphics[width=0.45\linewidth]{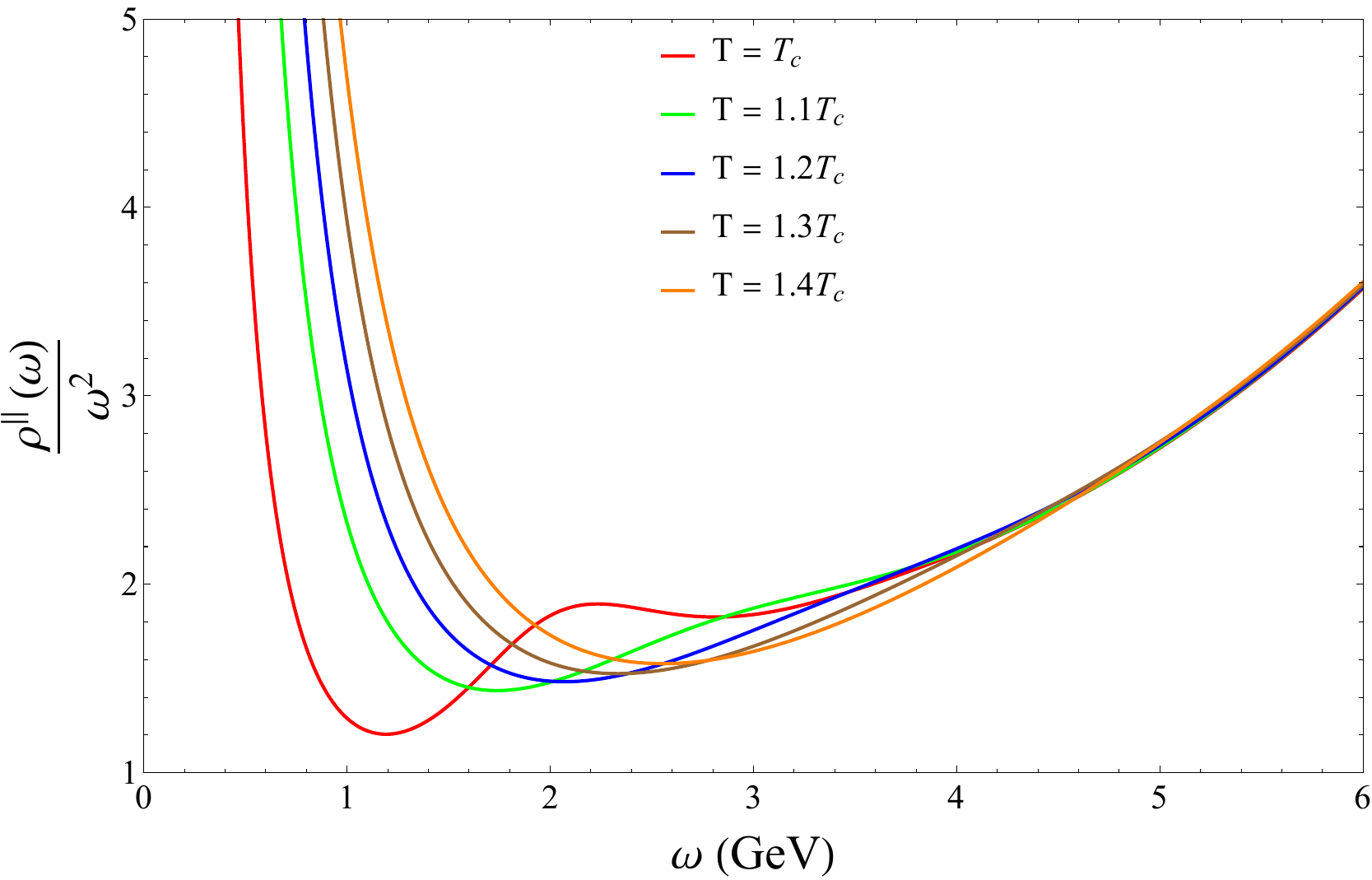}} 
  \label{spf_fixedB_parallelBPt2} \\
  \subfigure[$\pmb{B=0.4}$]{\includegraphics[width=0.45\linewidth]{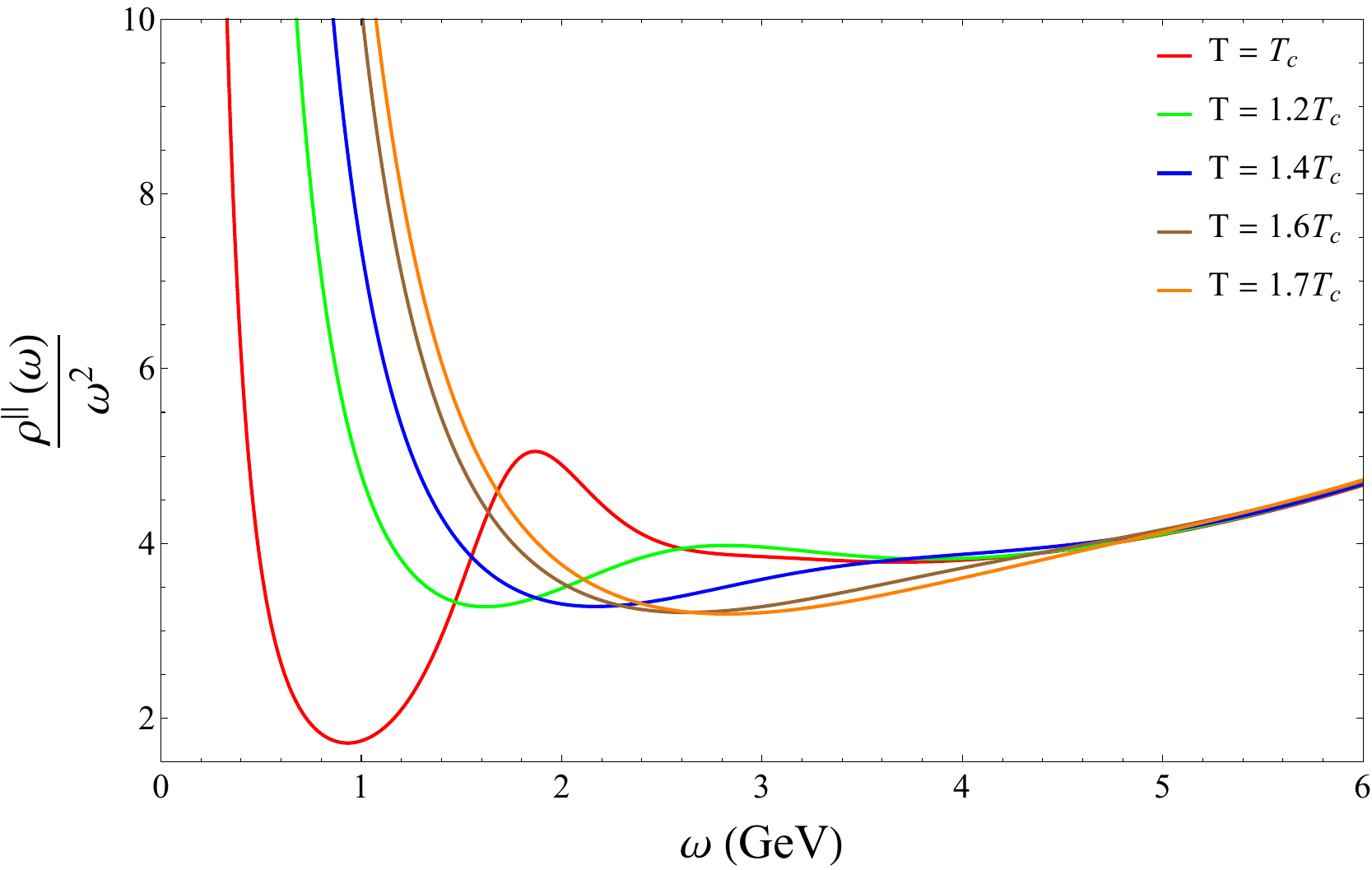}}
  \label{spf_fixedB_parallelBPt4}
  \subfigure[$\pmb{B=0.6}$]{\includegraphics[width=0.45\linewidth]{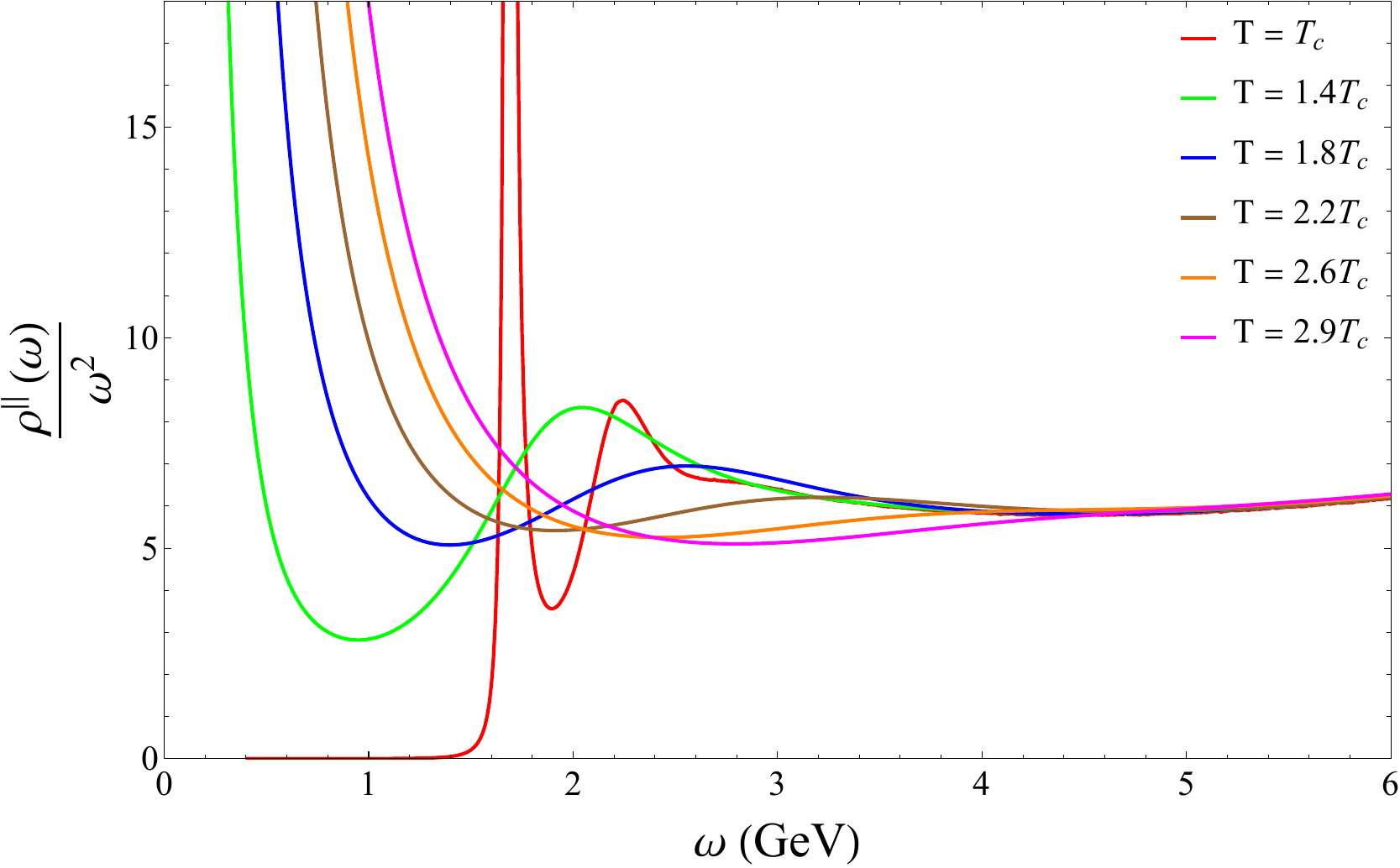}}
  \label{spf_fixedB_parallelBPt6}
  \caption{\small Variation of the spectral function with $\omega$ for varying temperatures with fixed $B$ in the parallel direction.}
  \label{spf_fixedB_parallel}
\end{figure}

However, we should note that $T_c$ depends non-trivially on $B$. For $B=0$, the critical temperature attains a value of $268~\text{MeV}$, while for $B=0.6$, $T_c$ decreases to $119~\text{MeV}$. Therefore, the critical temperature is substantially lower for higher magnetic field values. Accordingly, the broadening of the peaks for large $B$ values might be related to lower temperatures at which the bound quark states might still be present. To illustrate more on this, the variation of the spectral function for different temperatures and magnetic fields is shown in Fig.~\ref{spf_fixedB_parallel}. We see that for fixed $B$, there can be peaks at low temperatures, whose width smears out as the temperature increases and then completely disappears. This can be explicitly observed from Fig.~\ref{spf_fixedB_parallel}(c) where peaks can be observed till $T\sim 1.28~T_c=275~\text{MeV}$ for $B=0.4$, suggesting the existence of heavy quark bound state up to that temperature, whereas no such peak is observed above this temperature. Similarly, for $B=0.5$, peaks can appear even till $T\sim 1.6~T_c=286~\text{MeV}$. These results are consistent with lattice results where it was observed that the quarkonia could survive way above the deconfinement transition \cite{Asakawa:2003re}.

\begin{figure}[htb!]
	\centering
	\includegraphics[height=6cm,width=9cm]{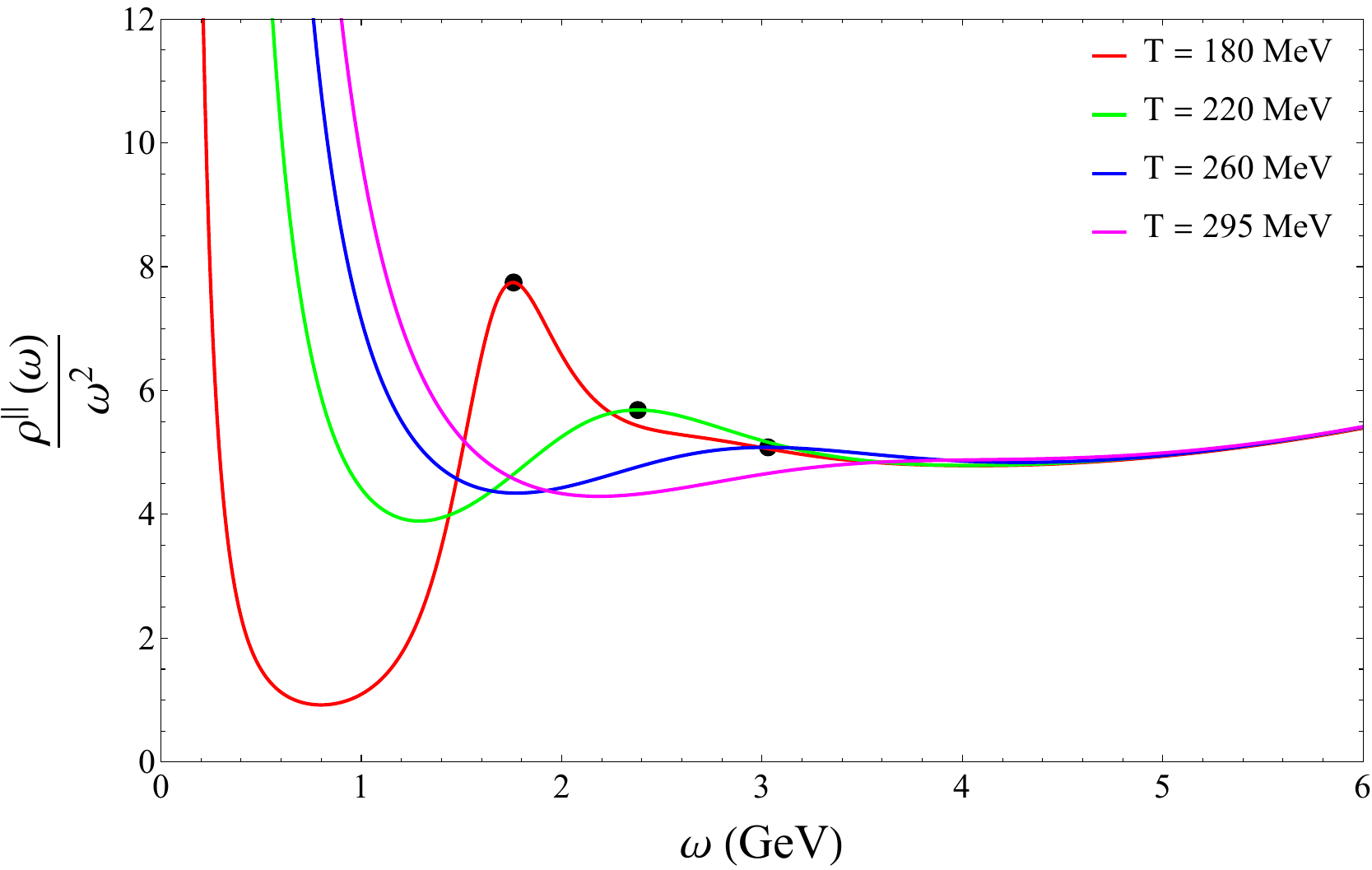}
	\caption{\small The vanishing of local maxima as an increase in temperature. }
	\label{localM}	
\end{figure}

To quantify the disappearance of peaks, we took the derivative of the spectral function with respect to $\omega$ and looked for the local maxima. The temperature at which the local maxima disappears accordingly gives the melting temperature. For $B=0.5~\text{GeV}$ and $b=0.044~\text{GeV}^2$, we show in Fig.~\ref{localM} the disappearance of the local maxima, seen as a black dot, as we increase the temperature to $T\sim 295~\text{MeV}$ (interpreted then as the melting temperature for this state). 

\begin{figure}[htb!]
	\centering
	\includegraphics[height=6cm,width=9cm]{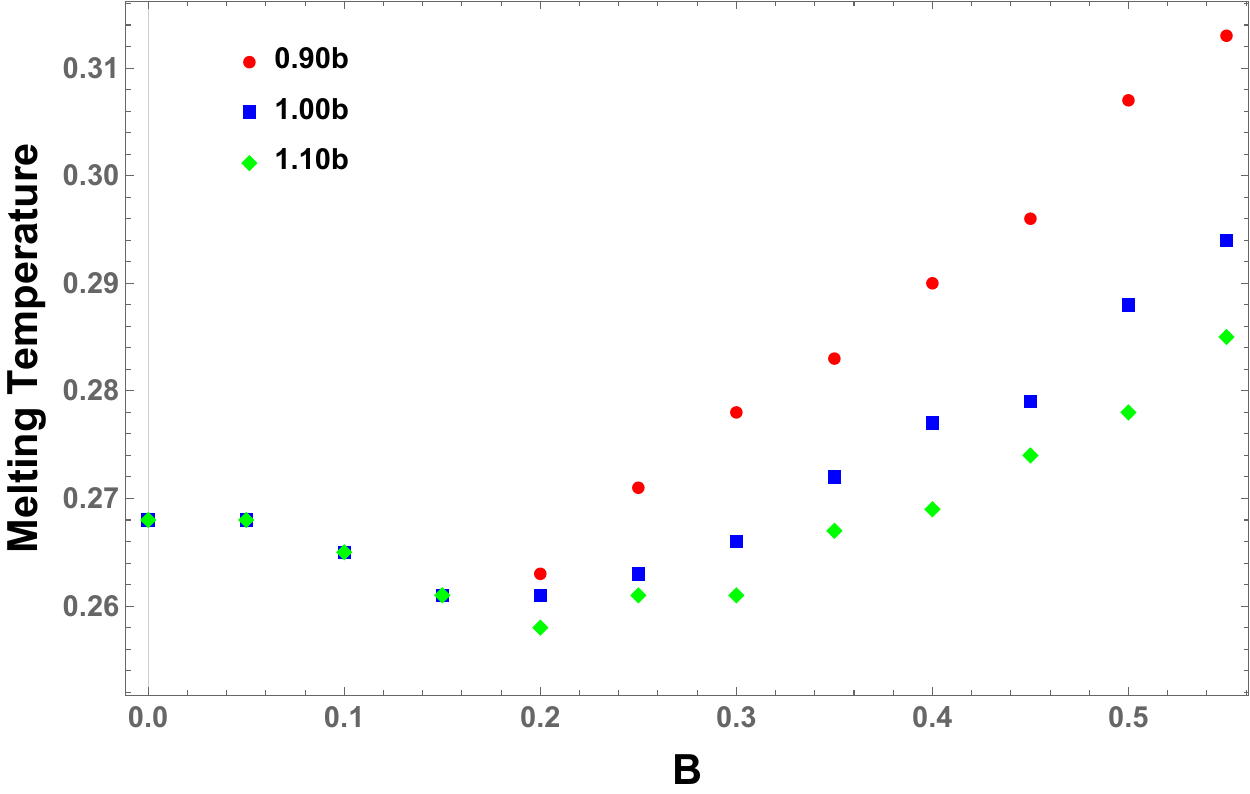}
	\caption{\small The variation of the melting temperature with magnetic field. }
	\label{meltingtempvsB}	
\end{figure}

We can also investigate how this melting temperature changes with the magnetic field. This is shown in Fig.~\ref{meltingtempvsB}. Our analysis suggests that the magnetic field has a non-trivial and intricate effect on the melting temperature. In particular, the melting temperature first decreases with the magnetic field and then increases with it. In Fig.~\ref{meltingtempvsB}, we have also presented results with a $10\%$ variation in $b$, corresponding to $10\%$ variation in the string tension. This certainly incorporates the error bar reported on the string tension from lattice QCD, cf.~the already mentioned $\sqrt{\sigma} =0.43\pm0.02~\text{GeV}$ of \cite{Bali:2000gf}.  We find that a switch from inverse magnetic to
magnetic catalysis in the melting temperature when the magnetic field increases is a generic phenomenon for all values of $b=0.044\pm10\%$. As of now, we do not have any lattice results to compare this interesting result to. This non-monotonic behavior of the melting temperature might be related to the non-monotonicity reported in the parallel/perpendicular lattice string tensions at higher magnetic field values, see \cite[Fig.~8]{DElia:2021tfb}. The ``stronger/weaker'' the confinement between heavy quarks, the larger/smaller the temperature we expect their bound states will melt at. However, since the polarization directions of the charmonia are not directly related, to the best of our knowledge, to the orientations of its constituent quark-antiquark pair (all relative to the magnetic field), this is only a loose intuition at best.

\subsubsection{Perpendicular case}
\begin{figure}[htb!]
	\centering
 \subfigure[$\pmb{T=T_c}$]{\includegraphics[width=0.45\linewidth]{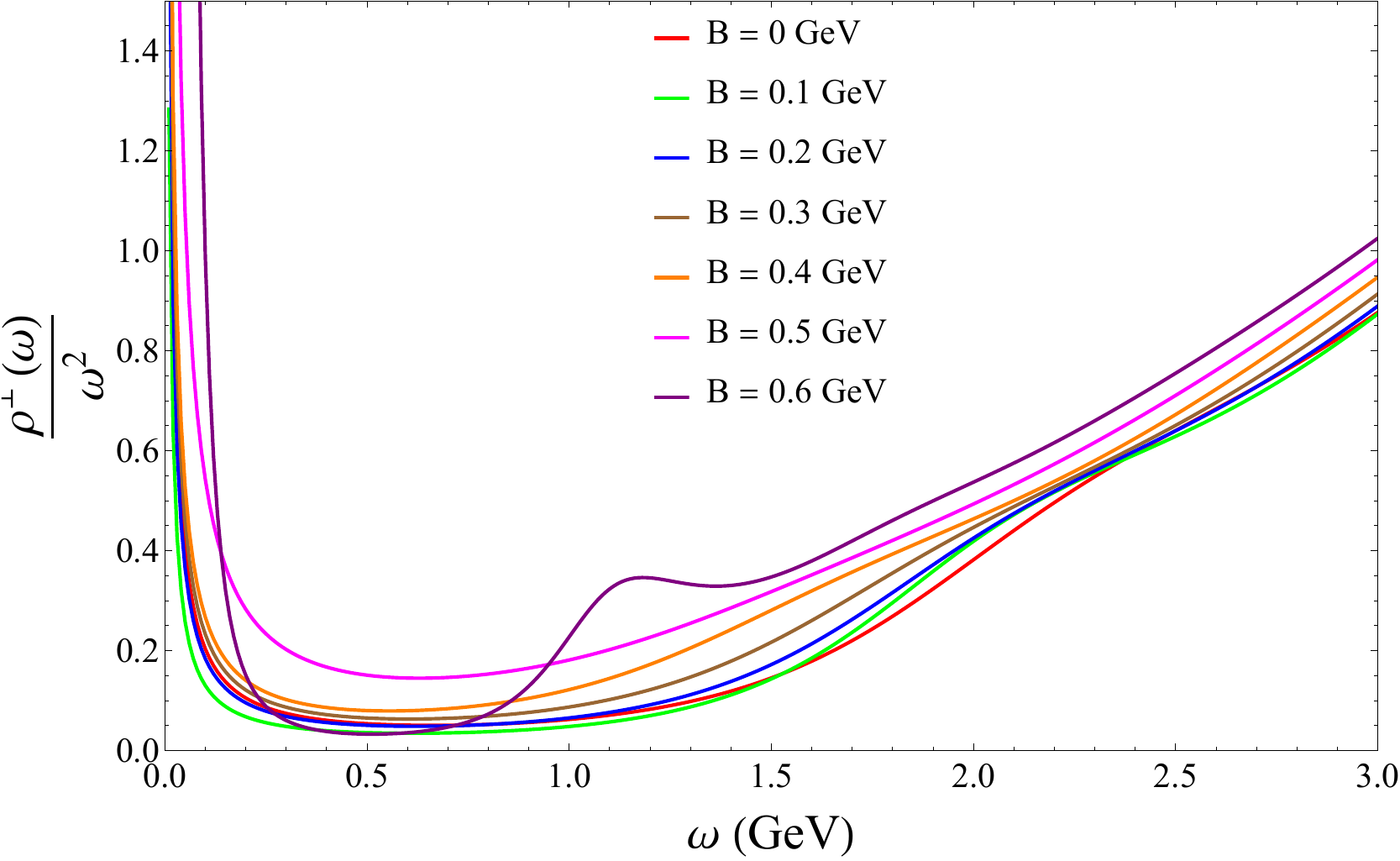}}
  \subfigure[$\pmb{T=1.5~T_c}$]{\includegraphics[width=0.45\linewidth]{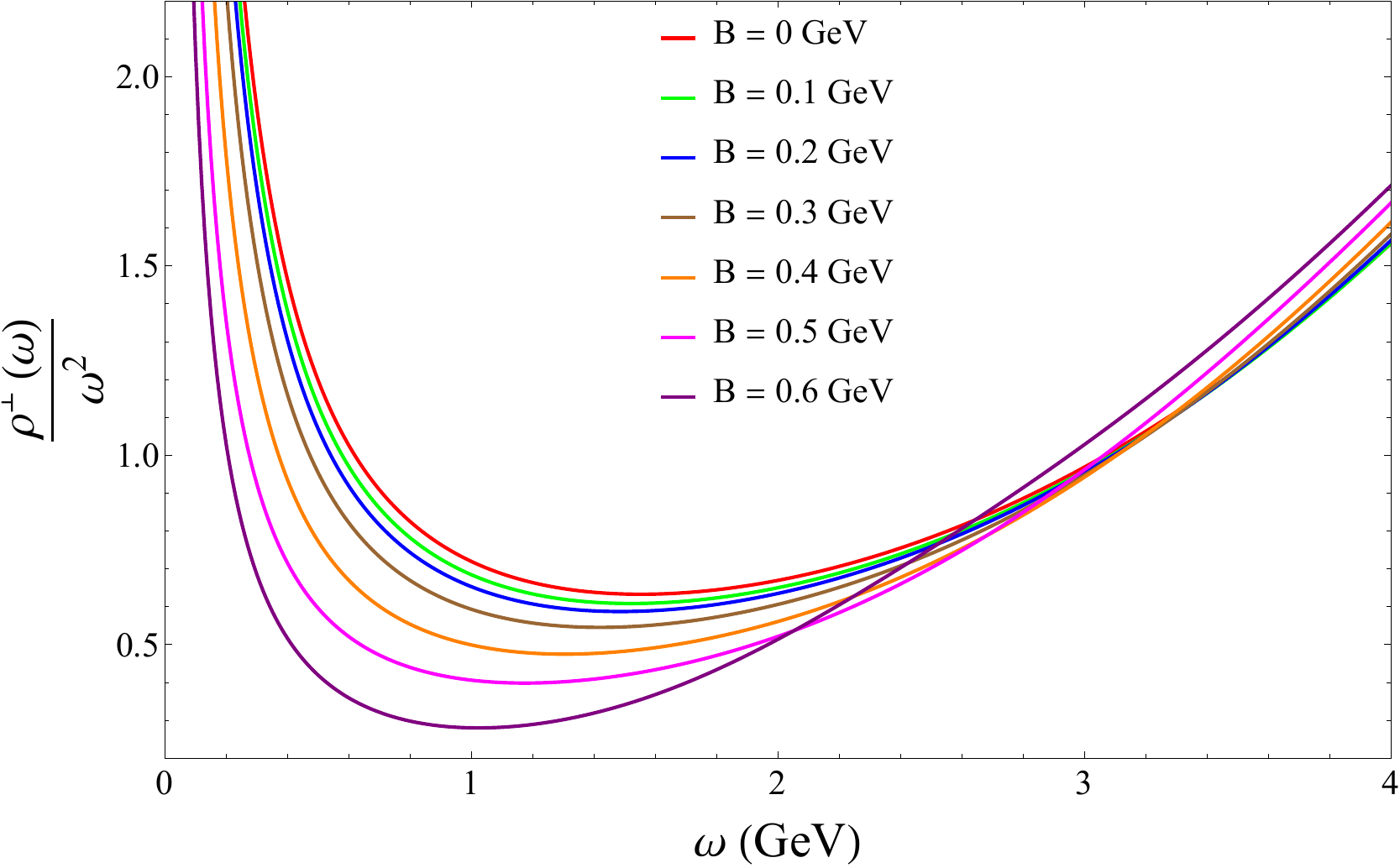}}
	\caption{\small Spectral function as a function of $\omega$ in the perpendicular direction for different magnetic fields. The temperature is fixed at $T=T_c$ (left) and $T=1.5~T_c$ (right). }
	\label{spf_diffB_perp}	
\end{figure}

\begin{figure}[htb!]
  \centering
  \subfigure[$\pmb{B=0.0}$]{\includegraphics[width=0.45\linewidth]{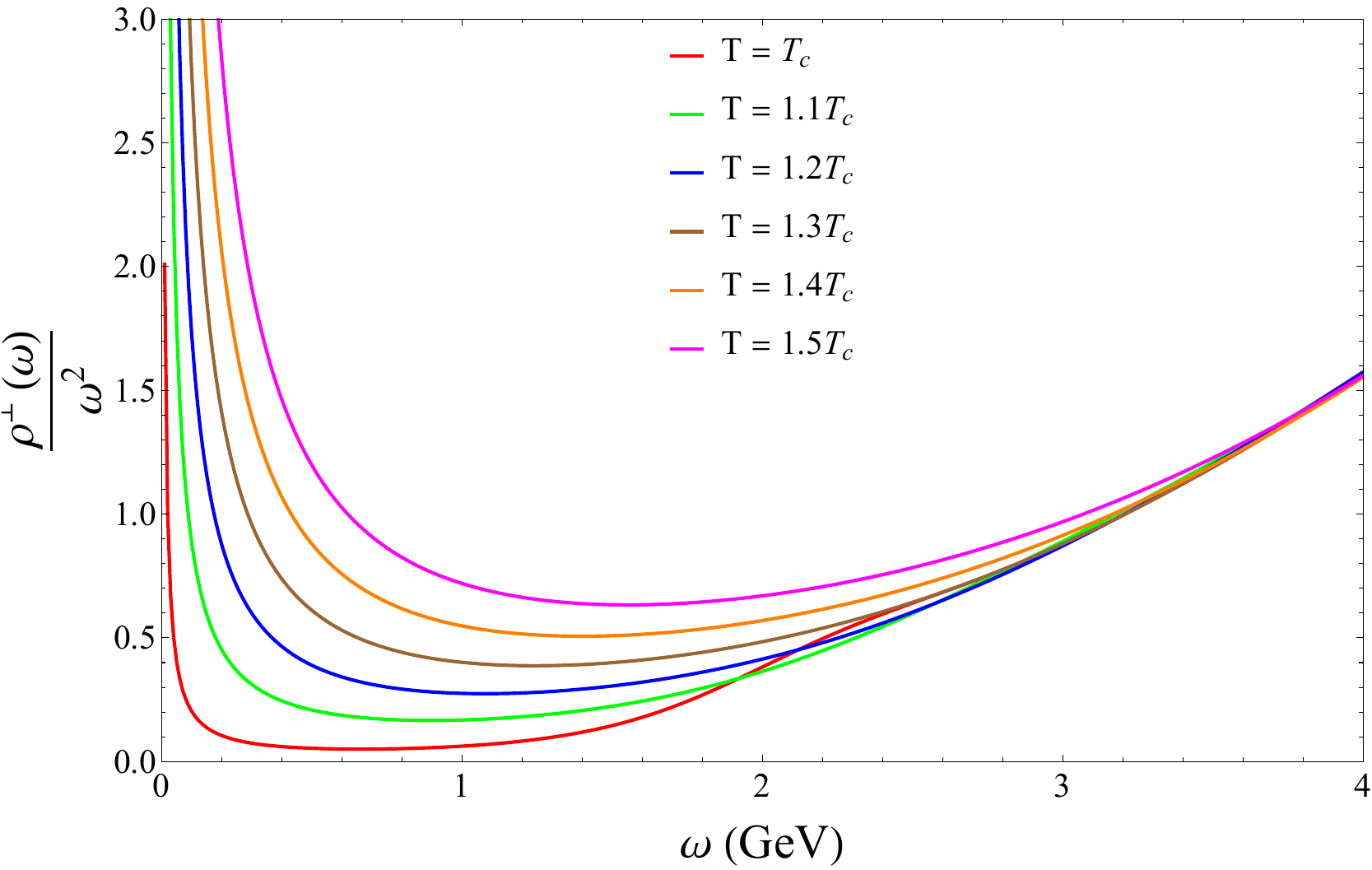}}
  \subfigure[$\pmb{B=0.2}$]{\includegraphics[width=0.45\linewidth]{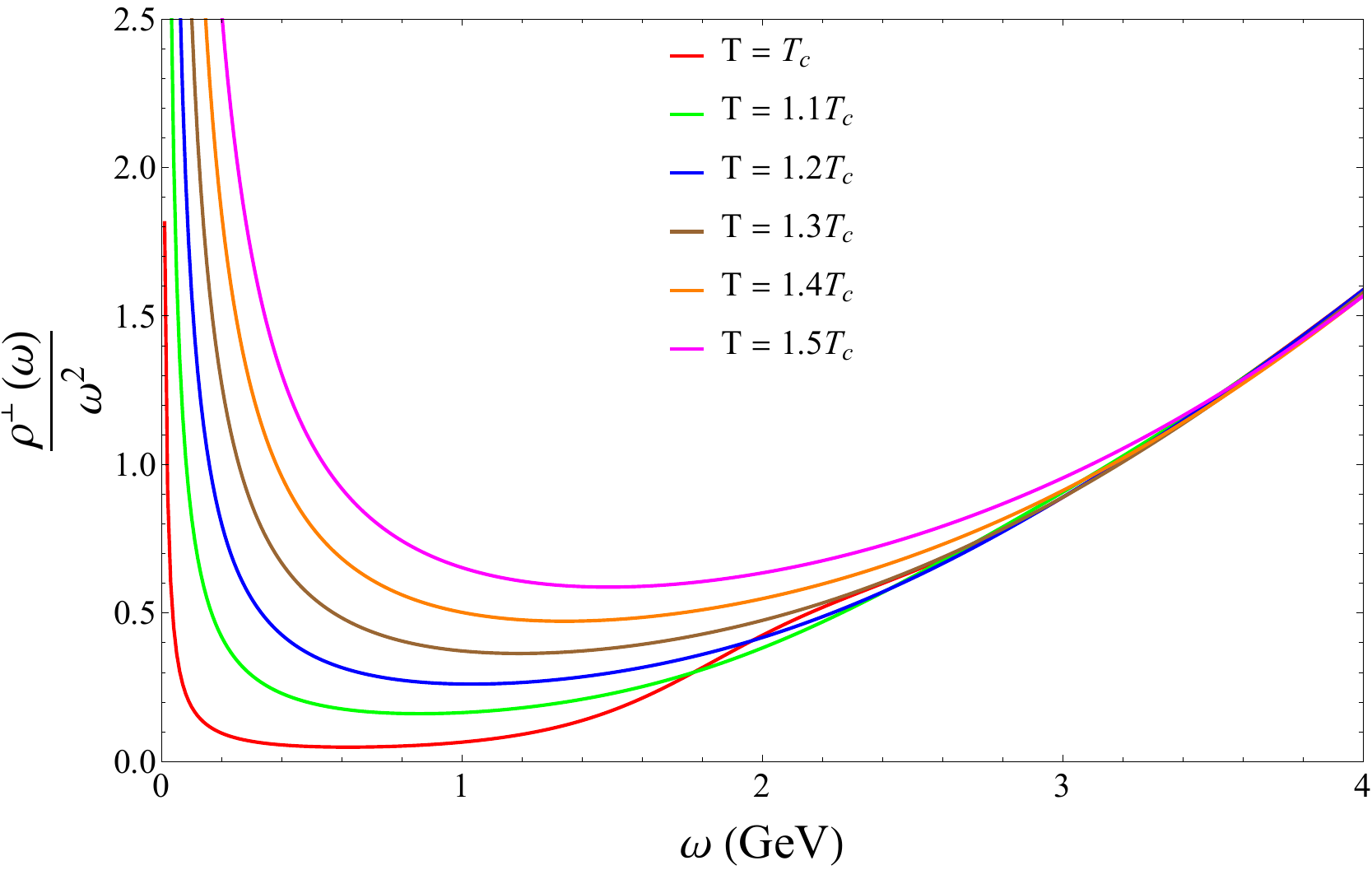}} \\
  \subfigure[$\pmb{B=0.4}$]{\includegraphics[width=0.45\linewidth]{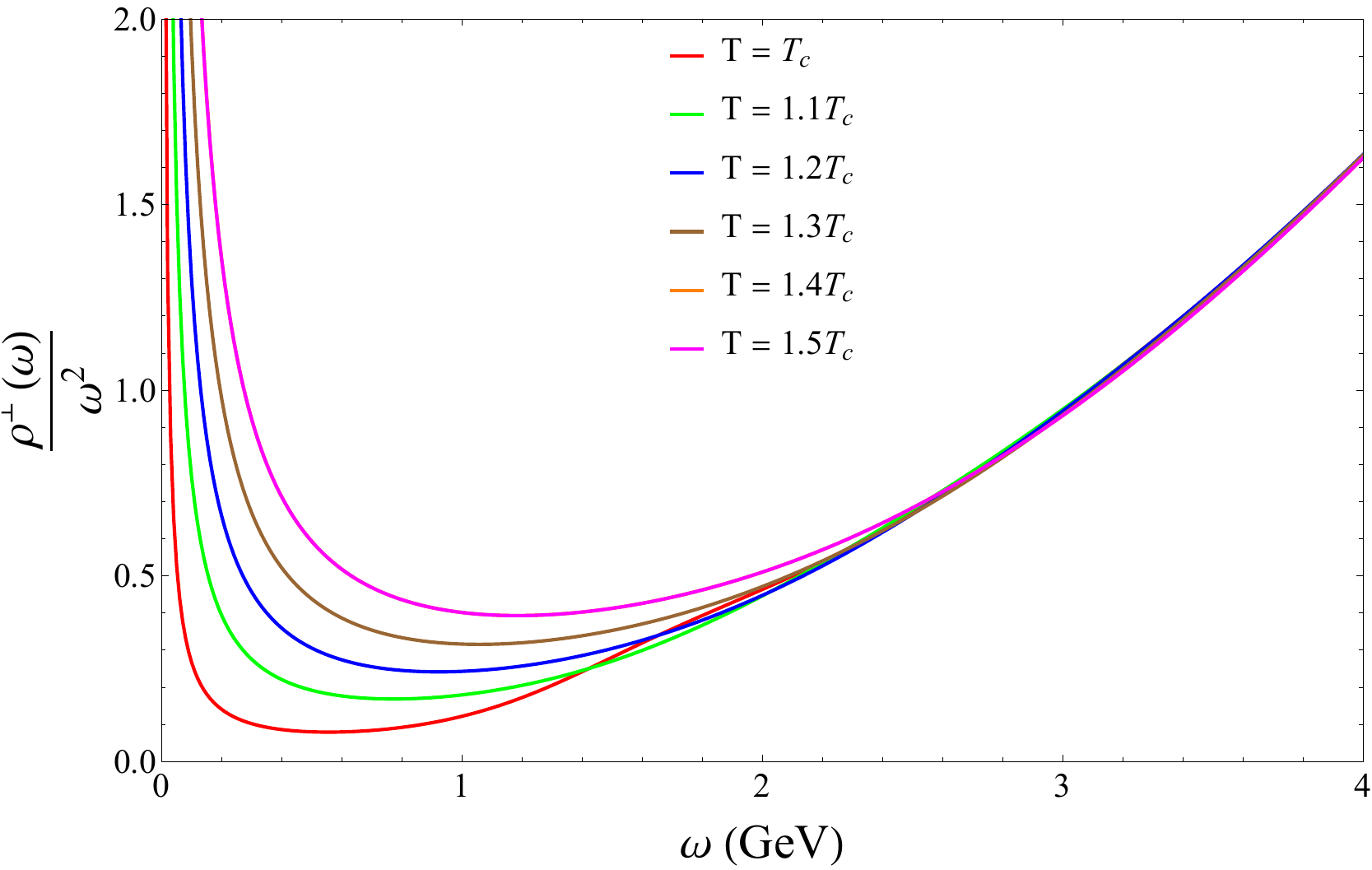}}
  \subfigure[$\pmb{B=0.6}$]{\includegraphics[width=0.45\linewidth]{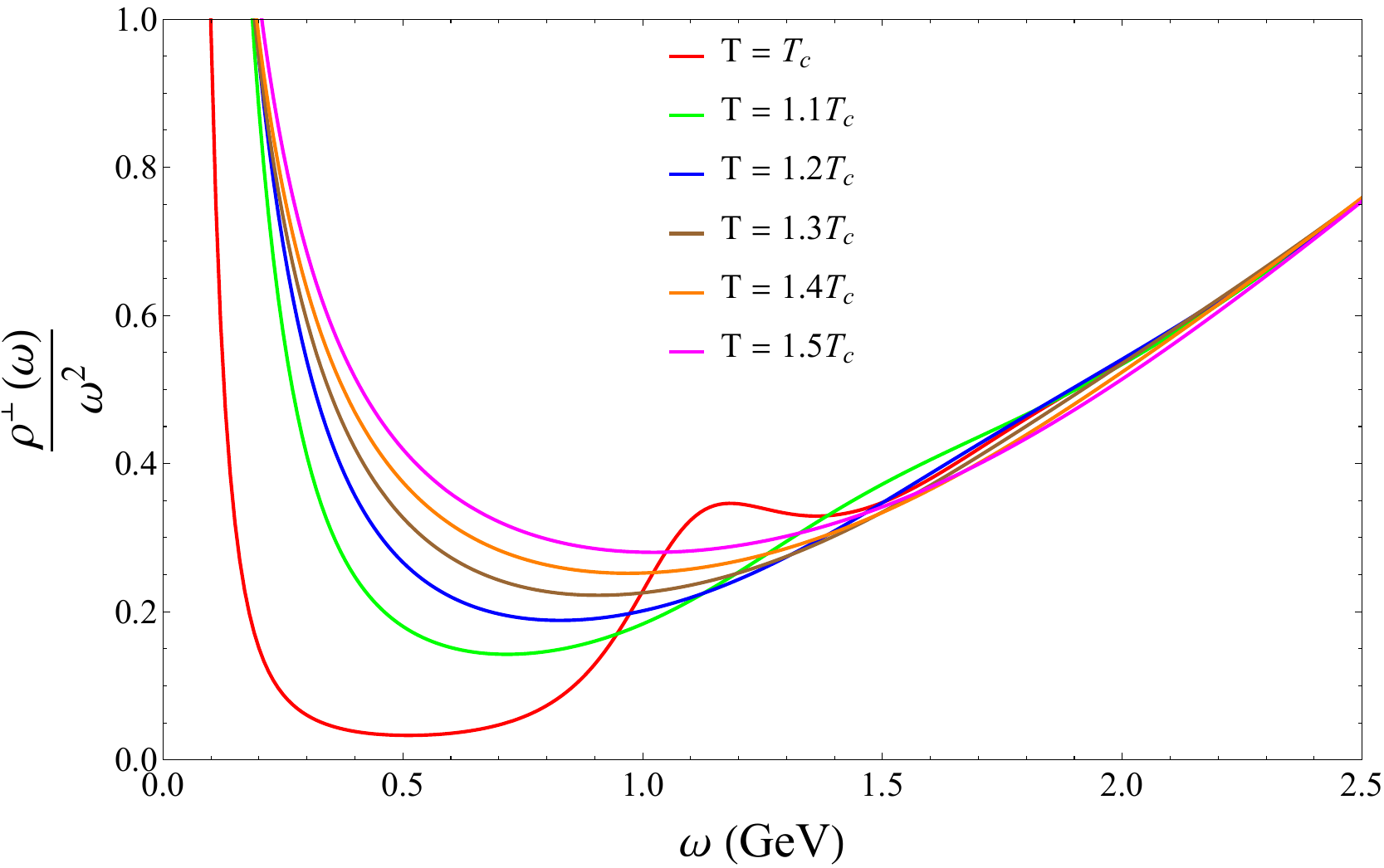}}
  \caption{\small Variation of the spectral function with $\omega$ for fixed values of $B$ with varying temperatures in the perpendicular direction.}
  \label{spf_fixedB_perpendicular}
\end{figure}

We now investigate the scenario where the polarization is perpendicular to the magnetic field. The behavior of the spectral function for different temperature and magnetic field values are shown in Fig.~\ref{spf_diffB_perp} and \ref{spf_fixedB_perpendicular}. The thermal behavior of the spectral peak remains similar to the parallel case when the temperature is varied. Specifically, the process of quarkonium melting unfolds with higher temperatures. This implies that, just like in the parallel case, higher temperatures play a catalytic role in promoting the melting phenomenon.


However, we also observe major differences compared to the parallel case. In particular, there are no peaks in the spectral function even at $T=T_c$ for magnetic field up to $B\sim0.5$, suggesting quarkonia melting even before the deconfinement sets in. This should be contrasted with the parallel magnetic field results where the peaks were present till $T=1.6~T_c$. The same is true for other magnetic field values as well. This implies that the quarkonia bound states can exist at relatively higher temperatures when the same magnetic field is applied in the parallel direction compared to the perpendicular direction.  At $T=1.1~T_c$, we do not observe any spectral peaks regardless of the strength of the magnetic field. This indicates at $T=1.1~T_c$, the quarkonium has melted regardless of whether there is a magnetic field or not. These phenomena signify the dissociation of quarkonia at elevated temperatures. Therefore, our overall holographic analysis suggests that, just like in the string tension, the magnetic field leaves anisotropic imprints in quarkonia melting, with more suppression in the perpendicular direction.

\section{Computation of the spectral function based on the membrane paradigm approach}
\label{sec7}
\subsection{Fluctuation equations of motion}
In this section, we use the membrane paradigm approach to recalculate the spectral function of heavy quarkonia \cite{Iqbal:2008by,BibEntry2024Jan}. This approach essentially states that at the linear response level, the low frequency limit of a strongly coupled boundary field theory at finite temperatures is governed by the horizon geometry of the corresponding dual black hole. Therefore, the generic boundary theory observables, such as the spectral function, can be expressed in terms of geometric quantities evaluated at the horizon.   Of course, the resulting spectral functions should be the same as those computed in the previous section, albeit obtained in a different way.

To obtain the spectral function using the membrane paradigm approach, we again start with the Born-Infeld part of the bulk action
\begin{equation}
    S=-b^2\int d^5x   f(\phi)\left(\sqrt{-g}-\sqrt{-\det\left(g_{\mu \nu}+ \frac{F_{\mu \nu}}{b}\right)}\right)\,.
    \label{fluctuationaction}
\end{equation}
Recalling from the previous section the gauge field equation of motion of the fluctuation can be recast as
\begin{equation}
    \partial_{\mu}\left(f(\phi)\sqrt{-\mathcal{G}} \tilde{F}^{\mu \nu}\right)=0\,,
    \label{mpeom}
\end{equation}
we can then associate a dual current with respect to a foliation by constant $z$-slices
\begin{equation}
    j^\mu=\sqrt{-\mathcal{G}}f(\phi)\tilde{F}^{z \mu}\,.
    \label{mpcurrent}
\end{equation}
The relevant equation of motion (\ref{mpeom}) can then be separated into two different directions: a parallel one involving fluctuations along ($t, x_1$) and a perpendicular one involving fluctuations along ($t, x_2~\text{or}~x_3$).

\subsubsection{Parallel case}
First, we will focus on the spectral peaks in the parallel case.
 Using Eq.~(\ref{mpcurrent}), the components $t$, $x_1$, and $z$ of Eq~(\ref{mpeom}) respectively give us,
\begin{eqnarray}
    -\partial_z j^t-\sqrt{-\mathcal{G}} f(\phi) \mathcal{G}^{tt} \mathcal{G}^{x_1 x_1} \partial_{x_{1} }\tilde{F}_{x_1 t}=0 \nonumber,\\
    -\partial_z j^{x_1}+\sqrt{-\mathcal{G}} f(\phi)\mathcal{G}^{tt} \mathcal{G}^{x_1 x_1} \partial_t \tilde{F}_{x_1 t}=0 \nonumber,\\
    \partial_t j^t+\partial_{x_1} j^{x_1}=0\,.
    \label{eomsmp}
\end{eqnarray}
From the Bianchi identity, we find the relation
\begin{eqnarray}
    \partial_z \tilde{F}_{x_1 t} -\frac{1}{\sqrt{-\mathcal{G}} f(\phi)} \mathcal{G}_{zz} \mathcal{G}_{x_1 x_1}\partial_tj^{x_1}-\frac{1}{\sqrt{-\mathcal{G}} f(\phi)} \mathcal{G}_{zz}\mathcal{G}_{tt}\partial_{x_1}j^t=0\,.
\label{bianchiidentity}
\end{eqnarray}
Now, we define the longitudinal conductivity $\sigma_L$
\begin{equation}
    \sigma_L(\omega,x_1,z)=\frac{j^{x_1}(\omega,x_1,z)}{\tilde{F}_{x_1 t}(\omega,x_1,z)}\,.
    \label{conductivity}
\end{equation}
The derivative of the above conductivity leads to
\begin{equation}
    \partial_z\sigma_L=\frac{\partial_z j^{x_1}}{\tilde{F}_{x_1 t}} -\frac{j^{x_1}}{\tilde{F}_{x_1 t}^2}\partial_z \tilde{F}_{x_1 t} \,.
    \label{conductivityderivative}
\end{equation}
Concentrating on a plane wave solution in the zero momentum limit, like in the previous section, and using Eqs.~(\ref{eomsmp})-(\ref{bianchiidentity}), we get
\begin{equation}
    \partial_z\sigma_L=i \omega \sqrt{\frac{\mathcal{G}_{zz}}{\mathcal{G}_{tt}}}\left[\frac{\sigma_L^2}{\zeta_L(z)}-\zeta_L(z)\right]\,,
    \label{PDFsigmaL}
\end{equation}
where,
\begin{equation}
    \zeta_L(z)=f(\phi)\sqrt{\frac{-\mathcal{G}}{\mathcal{G}_{zz}\mathcal{G}_{tt}}}\mathcal{G}^{x_1 x_1}\,.
\end{equation}
The differential equation of $\sigma_L$ can be numerically solved with the following initial condition
\begin{equation}
   \sigma_L(\omega,z_h)=\zeta_L(z_h) \,,
\end{equation}
corresponding to requiring regularity of $\sigma_L$ at the horizon. The above equation allows us to set up an initial value problem for Eq.~(\ref{PDFsigmaL}). One can therefore systematically obtain the conductivity $\sigma_L(\omega,0)$ at $z=0$ by numerically integrating from the horizon to the boundary. The spectral function is then obtained from the relation:
\begin{equation}
    \rho_\parallel (\omega)=\omega \text{Re} [\sigma_L(\omega,z=0)]\,.
\end{equation}
By employing linear response theory and Kubo's formula, it is also feasible to establish a connection between the conductivity, encompassing the spectral density, and the retarded Green's function
\begin{eqnarray}
    \sigma_L(\omega)=-\frac{G_R^L(\omega)}{i \omega}  \nonumber,\\
    \rho_\parallel(\omega)=- \Im [G_R(\omega)]\,.
\end{eqnarray}

The numerical results of the spectral function for the parallel case obtained from the membrane paradigm are numerically identical to those depicted in the last section, proving so that the methods are completely equivalent for various values of the magnetic field and temperature. 

\subsubsection{Perpendicular case}
Likewise, the evolution of the transverse channel is dictated by a dynamics equation\footnote{It is important to highlight that in our analysis, we are maintaining the momentum direction fixed along the $x_1$ direction. Consequently, when calculating the equation of motion for the transverse direction, we will consider components along $t$, $x_1$, $x_3$ and $z$.}
\begin{eqnarray}
     -\partial_z j^{x_3}-\sqrt{-\mathcal{G}} f(\phi)\mathcal{G}^{tt} \mathcal{G}^{x_3 x_3} \partial_t \tilde{F}_{t x_3}+\sqrt{-\mathcal{G}} f(\phi)\mathcal{G}^{x_1 x_1}\mathcal{G}^{x_3 x_3}\partial_{x_1}\tilde{F}_{x_1 x_3}= 0\,.
     \label{transverseeom}
\end{eqnarray}
Additionally, we have two constraints that are arising from the Bianchi identities \footnote{One arises from the components involving $t$, $x_3$, and $z$, while the other arises from the components involving $x_1$, $x_3$, and $z$.}
\begin{eqnarray}
  \partial_z \tilde{F}_{x_3 t} -\frac{1}{\sqrt{-\mathcal{G}} f(\phi)} \mathcal{G}_{zz} \mathcal{G}_{x_3 x_3} \partial_{t} j^{x_3}=0 \nonumber,\\
    \partial_{x_1} \tilde{F}_{t x_3 } + \partial_t \tilde{F}_{x_3 x_1 } =0\,.
    \label{perpendicularbianchi}
\end{eqnarray}
For the perpendicular case as well, we define the conductivity analogously
\begin{equation}
    \sigma_T(\omega,z)=\frac{j^{x_3}(\omega,z)}{\tilde{F}_{x_3 t}(\omega,z)}\,.
    \label{conductivitytransverse}
\end{equation}
Employing the identical methodology as in the longitudinal case one finds the following differential equation for $\sigma_T$
\begin{equation}
    \partial_z\sigma_T=i \omega \sqrt{\frac{\mathcal{G}_{zz}}{\mathcal{G}_{tt}}}\left[\frac{\sigma_T^2}{\zeta_T(z)}-\zeta_T(z)\right] \,,
    \label{PDEsigmaT}
\end{equation}
where,
\begin{equation}
    \zeta_T(z)=f(\phi)\sqrt{\frac{-\mathcal{G}}{\mathcal{G}_{zz}\mathcal{G}_{tt}}}\mathcal{G}^{x_3 x_3} \,.
\end{equation}
As in the parallel case, the spectral function is derived through an analogous procedure. The initial condition for solving the equation is determined by ensuring regularity at the horizon, i.e.,  $\sigma_T(\omega,z_h)=\zeta_T(z_h)$. The spectral function is then obtained from the relation:
\begin{equation}
    \rho_\perp (\omega)=\omega \text{Re} [\sigma_T(\omega,0)]\,.
\end{equation}



We can be rather short here, as our numerical results are again in full accordance with those of the previous section.

\section{Conclusion}
\label{sec8}
In this paper, we introduced a single flavour holographic Einstein-Born-Infeld-dilaton (EBID) model to effectively study the effects of an external magnetic field on quarkonia melting. In general, in Einstein-Maxwell-dilaton based holographic QCD models, Maxwell's equations describe the behavior of electric and magnetic fields linearly. In the context of quarkonium, the quarks inside the bound state have charges, but the overall quarkonium state is electrically neutral. Therefore, at a basic level, it seems reasonable that the electromagnetic fields would not directly affect the quarkonium state, as it lacks an overall charge to interact with the fields. However, quarkonium does possess an internal structure, including the distribution of charges within its constituent quarks, akin to an electric dipole moment. This internal structure could interact with external electromagnetic fields, leading to non-trivial effects on the quarkonium state, such as modifications to its dynamics or properties. To capture these effects, a more sophisticated model is needed. The EBID action provides a framework for describing the dynamics of charged objects in the presence of electromagnetic fields, accounting for the non-linearity and the internal structure of the object. In particular, the parameter $\ell_s \sqrt{q}$, with $q$ being the quark (or antiquark) charge, acts as a smearing charge parameter providing information about the overall charge distribution. Therefore, by extending the traditional Maxwell framework to incorporate EBI-like terms, one can hope to model better the interaction between quarkonium and external electromagnetic fields. In this work, we therefore constructed the first self-consistent dynamical magnetized holographic QCD model based on the EBID action. As a first comment, akin to what is observed in lattice QCD, our holographic model exhibits inverse magnetic catalysis for the deconfinement transition temperature.

We then employed the EBID model to explore quarkonium melting in the deconfinement phase. We investigated this using two different methods, viz.~the real-time holographic prescription and the membrane paradigm approach. Both approaches were leading to identical results nonetheless.
Our goal was to understand how temperature and magnetic field variations affect the melting process. We discovered several interesting features. Firstly, we observed that the quarkonium melting process sped up as temperature increased.  Additionally, our analysis suggested intriguing effects of the magnetic field on the melting. At higher magnetic field strengths, distinct spectral peaks emerged. This occurrence can be attributed to the intricate relation between the magnetic fields and the corresponding deconfinement transition temperature. Our analysis indicates that thermal effects exert a more pronounced influence on the melting than magnetic field effects. Moreover, we found a magnetic field related anisotropy in the melting spectrum. In particular, the bound state can exist at a relatively larger temperature, even above the deconfinement temperature, when the magnetic field is aligned in the longitudinal direction, in contrast with the transverse direction. Our analysis further indicated that the melting temperature is a non-monotonic function of the magnetic field, which first decreases and then increases.  

Moving forward, our focus will revolve around computing the transport characteristics of the QGP medium using this improved model. Primarily, we are interested in determining the magnetic field effects on the heavy quark number susceptibility, diffusion coefficients and their anisotropic behaviour. These quantities are important for understanding the medium's elliptic flow, given its inherent association with anisotropy, see e.g.~\cite{Fukushima:2015wck} and references therein.                                                                                                                                                                   
\section*{Acknowledgments}
\label{sec9}
 S.S.J.~ would like to thank Bhaskar Shukla and Arpan Bhattacharjee for their insightful discussions. The work of S.S.J.~is supported by Grant No. 09/983(0045)/2019-EMR-I from CSIR-HRDG, India. The work of S.M.~is supported by the core research grant from the Science and Engineering Research Board, a statutory body under the Department of Science and Technology, Government of India, under grant agreement number CRG/2023/007670. B.T.~thanks the São Paulo Research Foundation - FAPESP (Grants No. 2023/09097-9 and No. 2022/15325-1) for financial support. 
\appendix

\section{Equivalence of Born-Infeld actions}
\label{sec10}
Here, we show the equality of Eqs.~(\ref{BIlagrangian}) and (\ref{BIlagrangiansecond}). We depart from
\begin{equation}
	\mathcal{L}_{BI} = b^2\left(\sqrt{-g}-\sqrt{-\det\left(g_{\mu \nu}+ \frac{F_{\mu \nu}}{b}\right)}\right)\,.
	\label{BIlagrangianappend}
\end{equation}
For an antisymmetric matrix $X$, it holds that \cite{Alam2021Nov}
\begin{eqnarray}
\det(1+X)&=& 1- \frac{1}{2} Tr(X^2)+\det X \,\nonumber\\
	\det\left(g_{\mu \nu}+ \frac{F_{\mu \nu}}{b}\right)&= &\det\left[g_{\mu\rho} \left(\delta^\rho_\nu+\frac{F^\rho_\nu}{b}\right)\right]=g \det\left(\delta^\rho_\nu+\frac{F^\rho_\nu}{b}\right)\,,
\end{eqnarray}
where $g=\det(g_{\mu \nu})$. Let $X=\frac{F^\rho_\nu}{b}$. Then
\begin{eqnarray}
	\det\left(\delta^\rho_\nu+\frac{F^\rho_\nu}{b}\right) &=& 1+ \frac{1}{2 b^2} F_{\mu\nu}F^{\mu\nu}+\det\left(g^{\rho\mu \frac{ F^{\mu \nu}}{b}}\right) \,,  \nonumber\\
	&=& 1+  \frac{1}{2 b^2} F_{\mu\nu}F^{\mu\nu}+ \frac{1}{g} \det \left(\frac{ F^{\mu \nu}}{b}\right) \,.
	\label{equalityrelation}
\end{eqnarray}
The last term will be zero as $\det (F^{\mu \nu})=0$. Substituting (\ref{equalityrelation}) in the Lagrangian (\ref{BIlagrangianappend}), we get the desired result,
\begin{eqnarray}
	\mathcal{L}_{BI} &=& -b^2 \sqrt{-g -\frac{g}{2 b^2}F_{\mu \nu }F^{\mu \nu}} +b^2 \sqrt{-g} \,,  = b^2 \sqrt{-g} \left[1-\sqrt{1+ \frac{F_{\mu \nu} F^{\mu \nu}}{2 b^2}}\right]\,.
\end{eqnarray}

\bibliographystyle{JHEP}
\bibliography{mybib}

\end{document}